\documentclass[twocolumn]{aastex701}
\usepackage{amsmath}

\begin{document}

\title{Tidally Torn: Why the Most Common Stars May Lack Large, Habitable-Zone Moons}

\author[orcid=0000-0003-3920-7853,gname=Shaan,sname=Patel]{Shaan D. Patel}
\affiliation{Department of Physics, University of Texas at Arlington \\
Arlington, TX 76019, USA}
\email{shaan.patel@uta.edu}

\author[orcid=0000-0002-9644-8330,gname=Billy,sname=Quarles]{Billy Quarles}
\affiliation{Department of Physics and Astronomy, East Texas A\&M University \\ Commerce, TX 75428, USA}
\email{billylquarles@gmail.com}

\author[orcid=0000-0001-9194-2084,gname=Nevin,sname=Weinberg]{Nevin N. Weinberg}
\affiliation{Department of Physics, University of Texas at Arlington \\
Arlington, TX 76019, USA}
\email{nevin@uta.edu}

\author[orcid=0000-0002-8883-2930,gname=Manfred,sname=Cuntz]{Manfred Cuntz}
\affiliation{Department of Physics, University of Texas at Arlington \\
Arlington, TX 76019, USA}
\email{cuntz@uta.edu}

\begin{abstract}
Earth-like planets in the habitable zone (HZ) of M-dwarfs have recently been targeted in the search for exomoons.  We study the stability and lifetime of large (Luna-like) moons, accounting for the effects of 3-body interactions and tidal forces using the N-body simulator \texttt{rebound} and its extension library \texttt{reboundx}.  We find that those moons have a notably different likelihood of existence (and, by implication, observability).   Large moons orbiting Earth-like planets in the HZs of M4 and M2 dwarfs become unstable well before $10^7$ and $10^8 \textrm{ yr}$, respectively, and in most cases, those orbiting M0-dwarfs become unstable in much less than $10^9 \textrm{ yr}$. We conclude that HZ planets orbiting M-dwarfs are unlikely to harbor large moons, thus affecting the total number of possible moons in our galaxy and the Universe at large.  Since moons may help enhance the habitability of their host planet, besides being possibly habitable themselves, these results may have notable implications for exolife, and should also be considered when seeking solutions to the Drake equation and the Fermi paradox.
\end{abstract}

\keywords{exoplanets -- planets and satellites: dynamical evolution and stability -- tides -- habitable zone}

\section{Introduction} \label{sec:intro}

As the number of exomoon candidates increases, there is heightened interest in studies of exomoon formation, orbital stability, and habitability \citep{domingos06, heller14, Rosario-Franco2020}. Exomoon candidates include Kepler-1625b-i and 1708b-i \citep{teachey18,kipping22,kipping25}. In both cases the host stars are solar type \citep{mulders15}, although there is no a priori reason exomoons could not exist around other types of stars  \citep{williams97,kipping22}. Based on  detailed evaluations of both the initial mass function (IMF) and present-day mass function (PDMF) of the Milky Way as well as other galaxies \citep{kroupa01, kroupa02, chab03, wang25}, M-dwarfs are considered the most frequent type of stars.

In recent years, the investigation of exomoons orbiting HZ planets and specifically within M-dwarf systems has garnered increasing attention \citep{Kipping2013, hek4}, making it crucial to examine the conditions necessary for exomoons to exist in such systems.  Moreover, \cite{jwst24} proposed for (and was awarded) observations to search TOI-700d for a Luna-like moon analog orbiting the HZ terrestrial planet using the James Webb Space Telescope (JWST).
A particular aspect of exomoons concerns objects situated in stellar habitable zones (HZs).  Those are regions around a star at which liquid water could exist on a planetary or moon surface \citep{kasting93, kopp13, kasting14, ram18}.  HZs are also known as “Goldilocks zones,” where conditions might be just right for life based on the standard biochemical paradigm \citep{sullivan01}.  

Prominent scenarios of HZ exomoons include possible cases where such moons are hosted by Jupiter-type planets in distant HZs, potentially allowing habitability \citep{williams97,kalt17}, including HD 23079 \citep{tinney02}. Detailed dynamical studies identified regions of stability of HZ exomoons \citep{cuntz13, jagtap21, patel25}, including plausible tidal interactions within the system, which was found to be insufficient to induce significant outward migration toward the theoretical stability limit due to the large star--planet separation.
However, previous studies suggest significant dynamical challenges. \cite{kane17}, \cite{trif20}, and \cite{patel25b} found that strong perturbations and tidal effects generally prevent long-term moon survival in systems like TRAPPIST-1, GJ 1148, and K2-18, while \citet{marrod19}, using tidal evolution models, identified only a small subset of massive HZ planets where exomoons could remain stable for over 0.8 Gyr.

This study aims to assess the dynamical stability of exomoons within the HZs of M-dwarf stars through a comprehensive exploration of the parameter space for host planets with masses between 0.8 and 2.0 Earth masses, orbiting spectral types M0 to M4, while accounting for a range of tidal dissipation efficiencies\footnote{Contemporaneous with our work, \cite{Su2025} focus on warm planets with orbital periods of 10-200 days and the potential in-fall after planetary spin-down. They use a classical approach that neglects the mean motion perturbations potentially affecting the putative moon.}. Our methodology is based on a novel approach that enables us  to determine tidal migration timescales by including N-body integrations along with tidal effects. Simulations will be given for M0, M2, and M4 candidate stars.

This dynamical concept expands on previous purely analytical approaches by leveraging the latest features of the \texttt{reboundx} library, including recent expansions. Sets of model calculations are pursued for planets at different locations in M-dwarf HZs assuming they host Luna-size moons.  The orbital stability of those moons will be carefully assessed, which informs the kinds of systems that can be expected to exist in nature.

The Drake equation estimates the abundance of advanced civilizations in our galaxy by combining astrophysical and biological probabilities while the Fermi paradox highlights the contradiction between this expected abundance and the lack of evidence to date \citep{drake93, burchell06, kipping21, szabo24, kipping25b, civiletti25}. The possible scarcity of HZ moons around M-dwarfs could be a contributing factor to both.

Section 2 describes the methods used in this study, whereas Sect.~3 reports our results and discussion.
Our summary and conclusions are given in Sect.~4, whereas Sect.~5 conveys an outlook of future observational concepts allowing to test our theoretical predictions.

\section{Methods} \label{sec:methods}
\subsection{Numerical Simulation Setup}
We employ the N-body integrator \texttt{rebound} \texttt{(v4.4.3)} \citep{rebound}, in conjunction with the extension library \texttt{reboundx} \texttt{(v4.3.0)} \citep{reboundx}, to examine the orbital stability of exomoons surrounding Earth-like planets within the HZs of M-dwarf systems, incorporating tidal effects. The \texttt{tides\_spin} module \citep{tides_spin} uses a constant time-lag tidal model that follows the prescription of tides \citep{eggleton}. Our simulations use the \texttt{TRACE} integrator \citep{Lu2024} along with the \texttt{IAS15} integrator \citep{Rein2015} to model close encounters between the host planet and the potential moon, with the initial integration timestep set to 20\% of the moon's initial orbital period.

\subsection{Habitable Zone Limits} 
\begin{figure*}[ht!]
\plotone{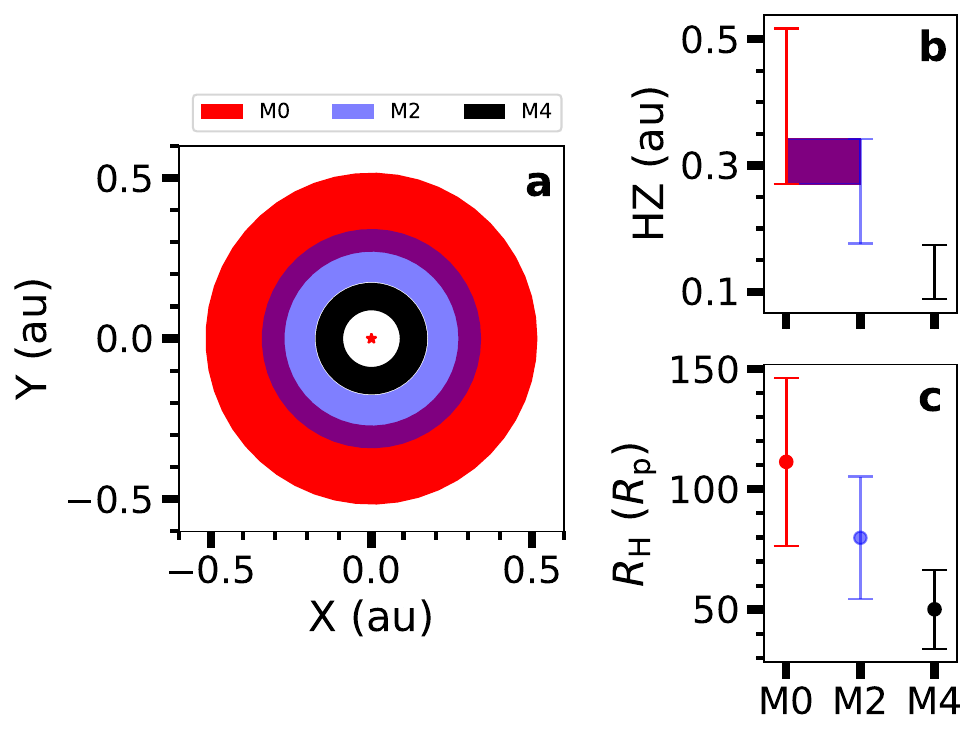}
\caption{HZs and Hill Radii of M0, M2, and M4 stars. (a) Top-down and (b) edge-on views of the HZs around M0 (red), M2 (blue), and M4 (black) dwarfs along with (c) the Hill radius for a prospective 1.0 $M_\oplus$ planet at the center of its respective HZ. The overlap of the M0 and M2 HZs is noted in purple in both (a) and (b).
\label{fig:hzrh}}
\end{figure*}

To determine the inner and outer HZ distances for specific M-dwarf spectral types (M0, M2, M4), we apply Equation 5 from \citet{Kopparapu_2014}. We adopt the 1.0 $M_{\oplus}$ constants for the runaway greenhouse (inner HZ) and maximum greenhouse (outer HZ) limits using the temperature and luminosity values reported in \citet{Pecaut_2013}; however, the selection of these constants is not expected to notably influence the resulting HZ boundaries. We use the inner and outer HZ distances as the range for our initial planetary semi-major axis values ($a_{\rm p}$) for an M0-dwarf ($0.27-0.52\ {\rm au}$), M2-dwarf ($0.17-0.35\ {\rm au}$), and M4-dwarf ($0.08-0.18\ {\rm au}$), with a step size of 0.001 au applied across all three cases; see Table~\ref{tab:init_con} and Fig.~\ref{fig:hzrh} for details.\footnote{
The computational cost of our longest running simulation is about 14 days, but the majority of the simulations finish within a 1-2 days where using a medium-large computing cluster makes the range of semi-major axis values tenable.}

\begin{deluxetable*}{lccccccccccccc}
\caption{Initial conditions for \texttt{rebound} simulations for M0, M2, and M4 systems.}
\label{tab:init_con}
\tablehead{\colhead{} & \colhead{$M_{\ast}$} & \colhead{$T_{\ast}$} & \colhead{$L_{\ast}$} & \colhead{$M_{\rm p}$}& \colhead{$a_{\rm p}$}& \colhead{$e_{\rm p}$}& \colhead{$k_{\rm 2}$}& \colhead{$\overline{C}$}
& \colhead{$P_{\rm s}$}& \colhead{$\tau_{\rm p}$}& \colhead{$M_{\rm m}$}
& \colhead{$a_{\rm m}$}& \colhead{$e_{\rm m}$}\\
\colhead{} & \colhead{($M_{\odot}$)} & \colhead{(K)} & \colhead{($L_{\odot}$)} & \colhead{($M_{\oplus}$)}& \colhead{(au)}& \colhead{}& \colhead{}& \colhead{} & \colhead{(h)}& \colhead{(s)}& \colhead{($M_{\oplus}$)} & \colhead{(au)}& \colhead{}}

\startdata
        M0 & 0.57 & 3850 & 0.0690 & 0.8--2.0 & 0.27--0.52 & 0.03 & 0.298 & 0.3308 & 5 & 698 & 0.0123 & $3R_{\rm Roche}$ & $10^{-6}$\\
	M2 & 0.44 & 3560 & 0.0290 & ... & 0.17--0.35 &... &... &... &... &... &... &... &... \\
        M4 & 0.23 & 3210 & 0.0072 & ... & 0.08--0.18 &... &... &... &... &... &... &... &... \\
\enddata
\tablecomments{The quantities $M_{\ast/\rm p/\rm m}$ represent the masses of the star, planet, and moon, respectively, while $a_{\rm p/\rm m}$ and $e_{\rm p/\rm m}$ denote the semi-major axes and eccentricities of the planet and moon. $R_{\rm Roche}$ represents the Roche limit of a fluid satellite. The symbols $T_{\ast}$ and $L_{\ast}$ correspond to the stellar surface temperature and luminosity. The parameter $k_{\rm 2}$ refers to the Love number of the planet, while $\overline{C}$, $P_{\rm s}$, and $\tau_{\rm p}$ represent the moment of inertia constant, the initial spin period, and the time-lag constant for the planet, respectively. Repeated parameters are indicated by ellipses.}
\end{deluxetable*}

\subsection{Initial and Stopping Conditions} 
We investigate Earth-like host planets with a planetary mass between 0.8 and 2.0 $M_{\oplus}$, stepping by 0.01 $M_{\oplus}$. The planetary radius is kept at a constant 1$R_\oplus$ for all runs while the host star and moon remain point masses. For comparison, the mean radius of TOI-700d is only slightly larger ($\lesssim 1.2 R_\oplus$) than an Earth-radius \citep{Gilbert2020, Gilbert2023}. The planet begins with an Earth-like Love number ($k_{2}= 0.298$), moment of inertia constant ($\overline{C} =0.3308$), and time-lag ($\tau_{\rm p} = 698$~s) \citep{neron97, bolmont15}, while the planetary spin period begins at 5 hours, which is consistent with our current knowledge for the spins of protoplanets \citep{Kokubo2010,Takaoka2023}. In all simulations, the mass of the moon ($M_{\rm m}$) equals that of Earth's moon, Luna, while the semi-major axis of the moon ($a_{\rm m}$) starts at $3\times$ the Roche limit of a fluid satellite, calculated by
\begin{align}
    R_{\rm Roche} \ = \ 2.44R_{\rm p}(\rho_{\oplus}/\rho_{\rm m})^{1/3},
\end{align}
where $\rho_{\oplus}$ and $\rho_{\rm m}$ denote the densities of the Earth ($5.515\ {\rm g~cm^{-3}}$) and Luna ($3.34\ {\rm g~cm^{-3}}$), respectively. See Table~\ref{tab:init_con} for additional information. Numerical simulations of the Giant Impact Hypothesis show that Earth's moon likely formed just exterior to $R_{\rm Roche}$ \citep{Canup2004}.  The fastest outward migration occurs when the tides are strongest (i.e., near $R_{\rm Roche}$) so that starting a moon from $1-3\ R_{\rm Roche}$ is somewhat arbitrary, where we make a fiducial choice. Each simulation evolves the star--planet--moon system up to 200 Myr, where a simulation is terminated if the moon's semi-major axis extends beyond the critical semi-major axis $a_{\rm crit}$, defined as 
\begin{align}\label{tab:stablim}
    a_{\rm crit} \ = \ 0.4031(1-1.123e_{\rm p})R_{\rm H},
\end{align}
where $e_{\rm p}$ and $R_{\rm H}$ are the eccentricity and the Hill radius of the planet, respectively \citep{Rosario-Franco2020}.

\citet{domingos06} provided a more optimistic criterion ($a_{\rm crit} \approx 0.4895\ R_{\rm H}$) due to a bias in their starting conditions \citep[see][]{Rosario-Franco2020}.  Hence, we opt for the more conservative result from \cite{Rosario-Franco2020} because their results uniformly sample the moon's starting orbital phase as our outwardly migrating moons encounter the chaotic region near the stability domain randomly. The moon's lifetime corresponds to the simulation termination time or the full simulation time ($200\ {\rm Myr}$). 

\subsection{Secular Tidal Evolution:} 
To explore longer evolution timescales of the moon, we utilize a secular tidal evolution  \citep{Barnes_2017} to compare to our \texttt{rebound} results. The following equations detail the semi-major axis, eccentricity, and spin axis evolution of both the moon and planet orbit with all symbols having their usual meaning:

\begin{align}
\frac{da}{dt} 
  &= \frac{2a^2}{G M_1 M_2} 
     \sum_{i=1}^2 Z_{\rm i} \bigg(
        \cos(\psi_{\rm i}) 
        \frac{f_2(e)}{\beta^{12}(e)} 
        \frac{\Omega_{\rm i}}{n} 
        - \frac{f_1(e)}{\beta^{15}(e)} 
     \bigg)
\end{align}

\begin{multline}
\frac{de}{dt} = \frac{11ae}{2GM_1M_2} \sum_{i=1}^{2} Z_{\rm i} \left( \cos(\psi_{\rm i}) \frac{f_4(e)}{\beta^{10}(e)} \frac{\Omega_{\rm i}}{n} - \frac{18}{11} \frac{f_3(e)}{\beta^{13}(e)} \right)
\end{multline}

\begin{multline}
\frac{\mathrm{d}\Omega_{\rm i}}{\mathrm{d}t} = \frac{Z_{\rm i}}{2M_{\rm i} r_{\rm g}^2 R_{\rm i}^2 n} \left( 2\cos(\psi_{\rm i}) \frac{f_2(e)}{\beta_{12}(e)} - \right.\\
\left.\left[1 + \cos^2 \psi_{\rm i} \right]\frac{f_5(e)}{\beta^9(e)}\frac{\Omega_{\rm i}}{n}\right)
\end{multline}
where
\begin{align}
Z_{\rm i} 
  &\equiv 3 G^2 k_{2,{\rm i}} M_{\rm j}^2 (M_{\rm i} + M_{\rm j})
     \frac{R_{\rm i}^5}{a^9} \, \tau_{\rm i} \label{eq:Zi}
\end{align}
and
\begin{equation}
    \begin{aligned}
        \beta(e) &= \sqrt{1 - e^2}, \\
        f_1(e) &= 1 + \frac{31}{2} e^2 + \frac{255}{8} e^4 + \frac{185}{16} e^6 + \frac{25}{64} e^8, \\
        f_2(e) &= 1 + \frac{15}{2} e^2 + \frac{45}{8} e^4 + \frac{15}{16} e^6, \\
        f_3(e) &= 1 + \frac{4}{3} e^2 + \frac{8}{9} e^4 + \frac{5}{64} e^6, \\
        f_4(e) &= 1 + \frac{3}{2} e^2 + \frac{1}{8} e^4, \\
        f_5(e) &= 1 + 3e^2 + \frac{3}{8} e^4.
    \end{aligned}
\end{equation}

These calculations explore expected moon lifetimes due to solely tidal effects in contrast with our numerical simulations which additionally account for mean motion and other non-secular effects. Each simulation evolves up to 5 Gyr utilizing the same termination criterion as in our numerical simulations with \texttt{rebound}. From these secular tidal theory calculations, we extend our \texttt{rebound} results for initial parameters $(a_{\rm p},\ m_{\rm p})$ that survive for 200 Myr while also investigating how this analytical method compares to our more robust numerical results.

\subsection{Maximum Lifetimes through Extrapolation} 
To estimate the maximum lifetimes of these long-lived cases (i.e., those that survive 200 Myr), we use an extrapolation method based on the \texttt{rebound} simulation output. After ${\sim}10\ {\rm Myr}$ the tidal migration proceeds quasi-linearly in logarithmic time, which allows us to apply a first-order regression to this late-stage migration. For each case, we randomly select 1,000 intervals with start times between 20–100 Myr and end times at the simulation limit. The moon’s stability limit is fitted (in log time), incorporating both $e_{\rm m}$ and $e_{\rm p}$ \citep{Rosario-Franco2020}, with an additional regression. The intersection of the migration and stability-limit fits defines the estimated time of instability, or maximum lifetime.

We test our method’s predictive power using simulations with lifetimes between 30–100 Myr as an ideal testbed. These cases work perfectly because we already know the final outcome (i.e., the moon's actual lifetime) and the migration period after 10 Myr is long enough to perform a reliable regression. We then compare our regression-based lifetime estimates against these known outcomes, computing the percent error to quantify any systematic bias in our method.

\section{Results and Discussion} \label{sec:results}

\subsection{General Comments}
Using \texttt{rebound} and \texttt{reboundx}, we perform N-body simulations with tidal effects to explore exomoon stability in a 3-body system around an M0, M2, and M4 dwarf. The mass $M_{\rm p}$ and semi-major axis $a_{\rm p}$ of the planet is varied with the $a_{\rm p}$ spanning the HZ of the given star. We track exomoon lifetime up to 200 Myr following the moon's formation, where individual simulations are terminated when the moon crosses the stability limit \citep{Rosario-Franco2020}. 

\begin{figure*}[ht!]
\plotone{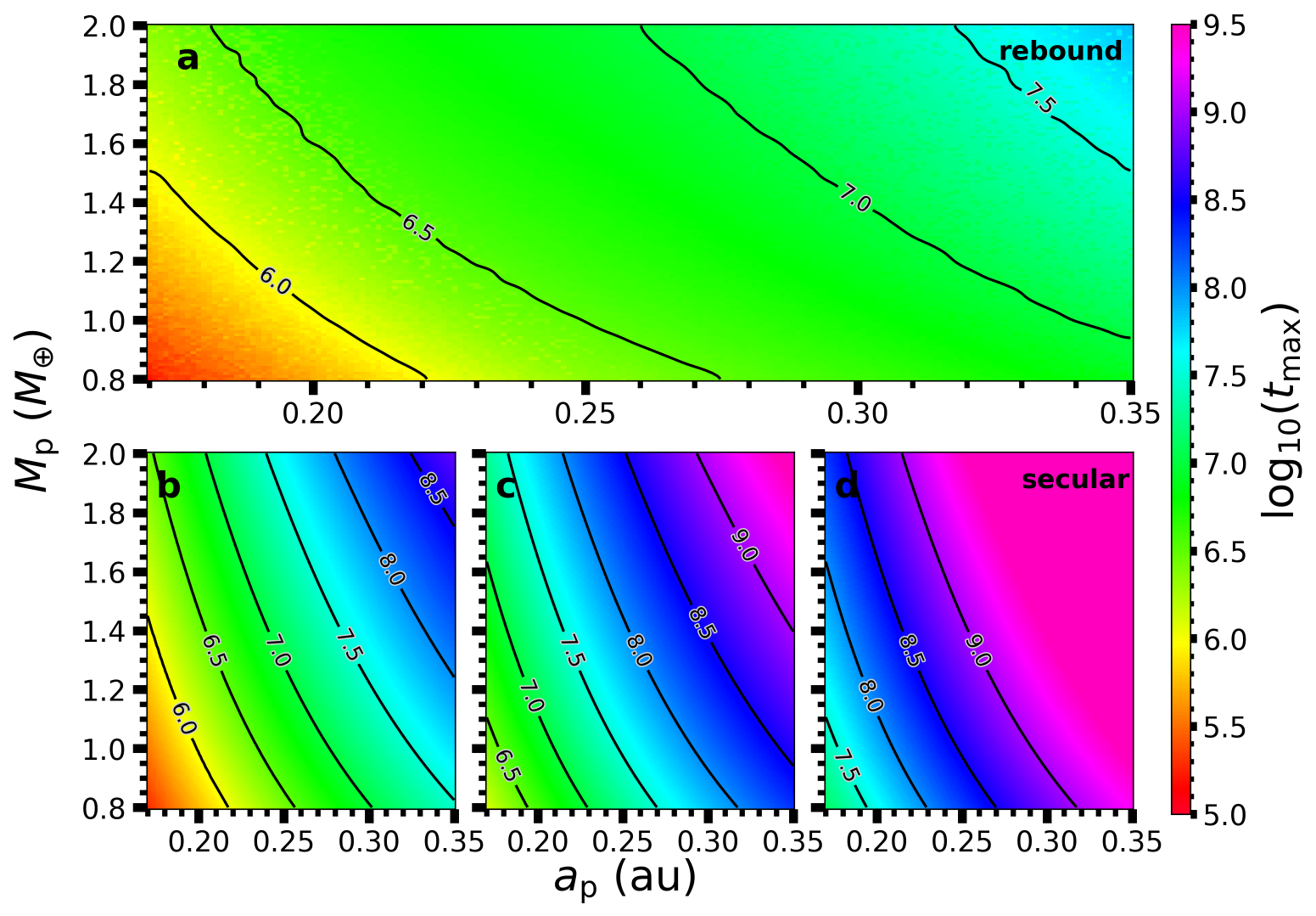}
\caption{Stability simulation results for a HZ planet--moon orbiting an M2-dwarf. Logarithm of maximum moon lifetime ($\log_{10}\left[t_{\rm max}\right]$) from simulations considering a range of host planet masses (0.8 to 2 $M_\oplus$) and semi-major axis within the HZ of an M2-dwarf. (a) shows results from N-body simulations using \texttt{rebound} with a tidal time-lag $\tau_{\rm p}$ of 698 s, while (b-d) show outcomes from secular tidal theory \citep{Barnes_2017} with tidal dissipation timescales $\tau_{\rm p}$ of (b) 698 s, (c) 100 s, and (d) 10 s. Contour lines indicate $\log_{10}(t_{\rm max})$ values, reflecting the time of moon loss due to tidal migration. Red, yellow, green, and blue regions represent increasingly longer lifetimes while purple regions represent the longest lifetimes.
\label{fig:m2_param}}
\end{figure*}

By varying the planet’s mass and its semi-major axis, we broadly sample the region where the planet’s gravity dominates because both parameters can modify the planet’s Hill radius. Broadly, our numerical simulations show that moon lifetimes scale with the Hill radius: a larger Hill sphere lengthens the migration path before escape. The host star and the moon tidally interact with the planet to slow down its spin.  Since the moon's orbit initially lies beyond the co-rotation radius, the moon migrates outward.  This tidal interaction is modulated by the tidal time-lag $\tau_{\rm p}$, which sets the phase offset between the tidal bulge and the line of centers.

The tidal time-lag of the planet quantifies how efficiently tidal energy is dissipated and the angular momentum transferred to the moon via a torque. Varying $\tau_{\rm p}$ reflects how composition and structure shape the tidal response through de-spinning, and the moon's orbital evolution. The modern Earth has a $\tau_{\rm p}$ of 698 seconds, which we adopt in most simulations and represents the tidal strength expected for a world with a significant water ocean.  In contrast, we also employ models with weaker tides ($\tau_{\rm p}=100$ and $10$ seconds) using secular tidal theory \citep{Barnes_2017}, which evaluate an expected dry-Earth scenario and an even more extreme (yet unknown) composition, respectively. In the absence of information on the true composition of these worlds, we characterize the dry-Earth as a ``weak tide" and the case of $\tau_{\rm p}=10\ {\rm s}$ as a ``very weak tide". The tidal dissipation efficiency of a planet can also be expressed via the tidal quality factor, $Q_{\rm p}$ \citep[see recent examples from][]{Leconte2010,heller11,Barnes_2017}.

According to \citet{sasaki12}, $Q_{\rm p}$ typically ranges from 10--500 for rocky planets and exceeds $10^4$ for ice and gas giants. Using the $Q_{\rm p}$--$\tau_{\rm p}$ relation ($\tau_{\rm p} \sim 1/Q_{\rm p}$) \citep{Leconte2010,heller11}, a time-lag of $\tau_{\rm p}=100$ s corresponds to $Q_{\rm p} \approx 638$, near the upper limit for rocky bodies, while $\tau_{\rm p}=10$ s gives $Q_{\rm p} \approx 6380$, approaching values typical of Jovian planets. These cases represent weaker tidal dissipation regimes, serving as points of comparison to the Earth-like baseline. Secular tidal theory allows us to probe billion-year timescales where N-body simulations are too computationally expensive.

\begin{figure*}[ht!]
\plotone{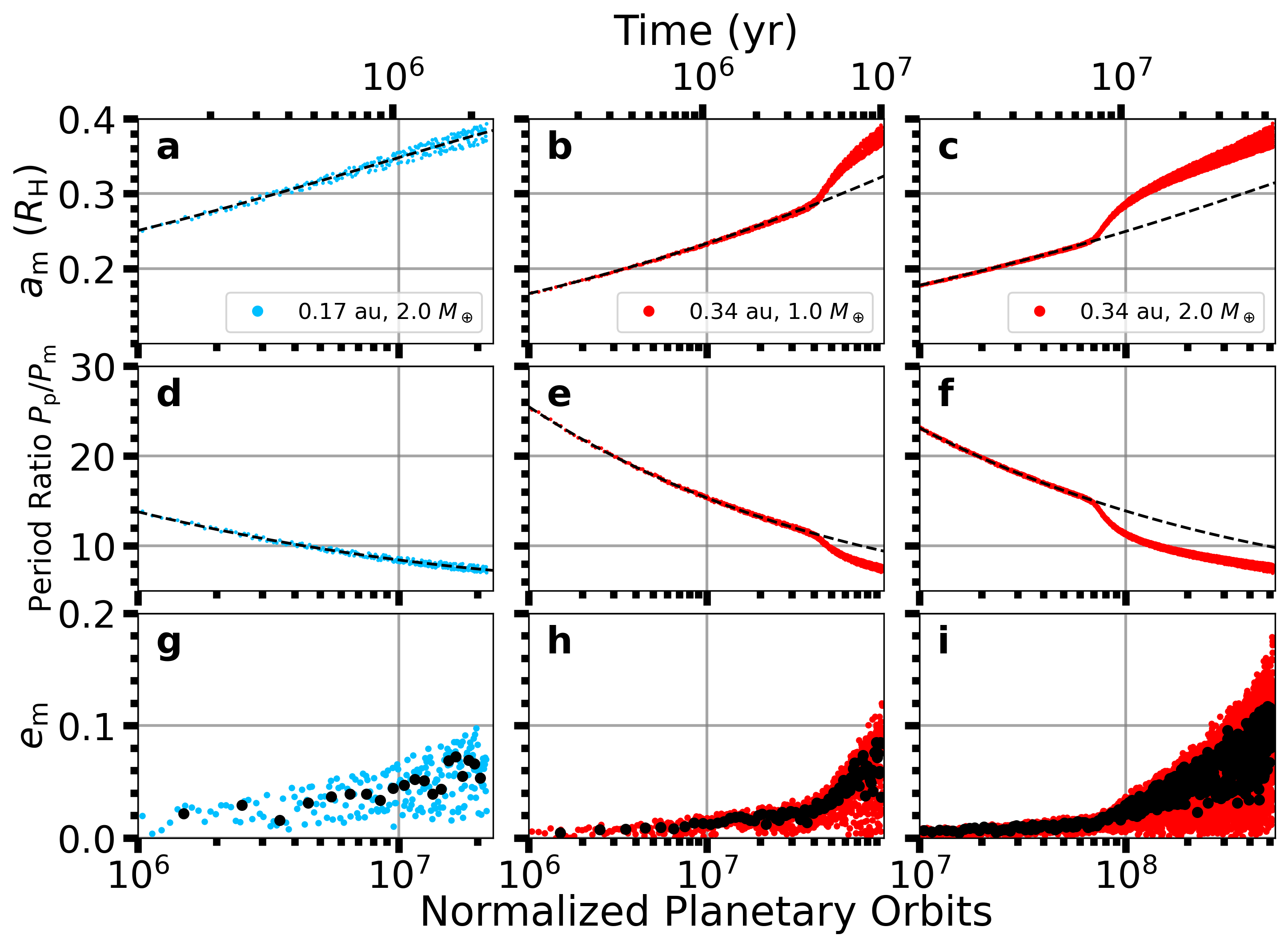}
\caption{Time evolution for three HZ planet--moon systems orbiting an M2-dwarf. Time evolution of the (a-c) exomoon's semi-major axis $a_{\rm m}$, (d-f) planet to moon period ratio $P_{\rm p}/P_{\rm m}$, and (g-i) exomoon's eccentricity $e_{\rm m}$ for 3 distinctive points in the parameter space (0.17 au and 2.0 $M_\oplus$, 0.34 au and 1.0 $M_\oplus$, 0.34 au and 2.0 $M_\oplus$) in the M2 case. The bottom $x$-axis, plotted on a logarithmic scale, shows time normalized by the orbital period of a planet at the inner edge of the HZ (0.17 au) while the top $x$-axis shows time (in years). The blue and red points signify \texttt{rebound} results while black dashed lines represent results from secular tidal theory. Black dots in the eccentricity panels represent the median value over 1 million orbits.
\label{fig:m2_time}}
\end{figure*}

\subsection{Habitable Zone Moons in M2-dwarf systems} \label{sec:m2_systems}
Figure \ref{fig:m2_param} illustrates the results of our simulations with color-coded cells where the color represents the exomoon's lifetime depending on the initial condition ($a_{\rm p}$, $M_{\rm p}$).  With Earth-like tides ($\tau_{\rm p} = 698\ {\rm s}$), the \texttt{rebound} results (Fig.~\ref{fig:m2_param}a) differ markedly with those using the tidal secular approximation (Fig. \ref{fig:m2_param}b), where moon lifetimes are $<100\ {\rm Myr}$ for the former and up to ${\sim}100\ {\rm Myr}$ for the latter. When the moon resides deep within the planet’s potential well, the secular tidal equations predominantly govern its evolution and the N-body perturbations are minimized. Eventually, mean motion resonances (MMRs) and other three-body effects become significant leading to the observed divergence in evolutionary behavior. Since \texttt{rebound} accounts for the increasingly significant perturbations on the mean-motion timescale, its results are considered more physically reliable than those derived from the tidal secular theory, which assumes that these perturbations are always small. If we assume weaker tides ($\tau_{\rm p} = 100\ {\rm s}$; Fig.~\ref{fig:m2_param}c), the moon lifetimes extend to ${\sim}1-2$ Gyr. In the very weak regime ($\tau_{\rm p} = 10\ {\rm s}$; Fig.~\ref{fig:m2_param}d), some moons survive nearly 5 Gyr before reaching the stability limit.

From the \texttt{rebound} results, we find that the moon migrates to the stability limit between $\sim 1-50$ Myr following a two-phase migration. The first phase of evolution aligns with the secular tidal model while the second phase is triggered as the moon crosses the $15:1$ MMR.  As the moon migrates outward, each resonance crossing takes longer and imparts a stronger kick, amplifying the moon's eccentricity (Figs.~\ref{fig:m2_time}a-f). Due to the slower migration, the MMRs force the orbital evolution to deviate from the secular tidal model. In this second phase, the moon spends more time within the MMRs and the kicks from the host star lead to higher $e_{\rm m}$ that are not damped by tidal dissipation (Figs.~\ref{fig:m2_time}g-i).

\begin{figure*}[ht!]
\plotone{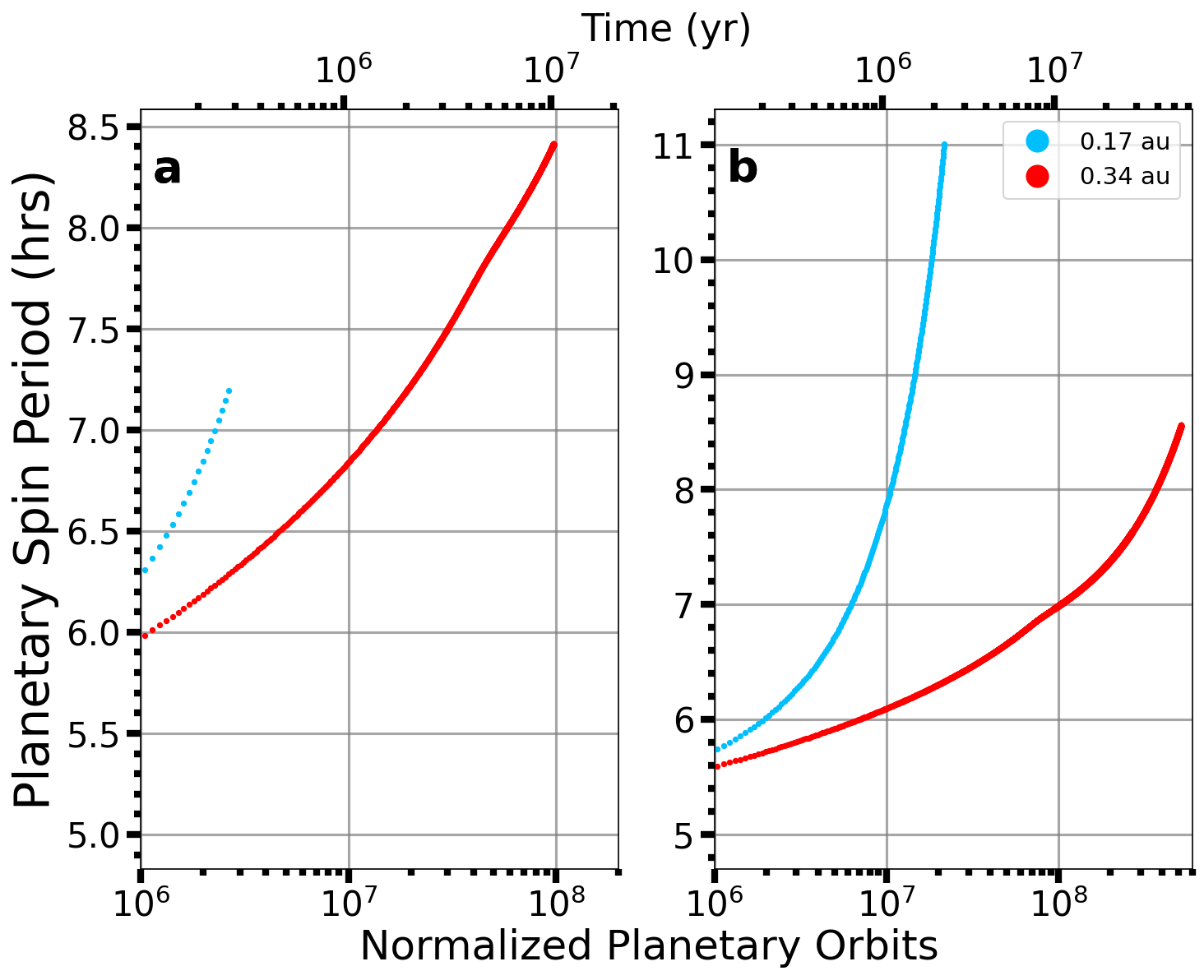}
\caption{Spin evolution for four HZ planet--moon systems orbiting an M2-dwarf. Evolution of planetary spin period for two $a_{\rm p}$ cases in the M2 parameter space (0.17 au, 0.34 au) at (a) 1.0 $M_\oplus$ and (b) 2.0 $M_\oplus$ from \texttt{rebound} simulations where curve lengths represent lifetimes. The bottom $x$-axis, plotted on a logarithmic scale, shows time normalized by the orbital period of a planet at the inner edge of the HZ (0.17 au) while the top $x$-axis shows time (in years).
\label{fig:m2_spin}}
\end{figure*}

If we start the planet at the inner edge of the HZ ($a_{\rm p} = 0.17\ {\rm au}$; Figs. \ref{fig:m2_time}a, \ref{fig:m2_time}d, and \ref{fig:m2_time}g), the \texttt{rebound} evolution generally follows the secular tidal evolution. Due to the faster spin-down of the planet from the star and the quicker overall evolution of the system, the tidal migration overwhelms the 3-body and MMR effects. The moon’s eccentricity still increases because the secular forcing timescale is only ${\sim}$ $10^3-10^4$ yr, which is short compared to the migration timescale.

If we instead start the planet at the outer edge of the HZ ($a_{\rm p} = 0.34\ {\rm au}$; Figs. \ref{fig:m2_time}c, \ref{fig:m2_time}f, and \ref{fig:m2_time}i), the \texttt{rebound} simulations only follow the secular tidal evolution up to $\sim 5-8$ Myr, after which the second phase in the evolution occurs due to MMR exposure and the rebound simulations deviate from the secular calculations. After the bump in the $a_{\rm m}$ evolution, the migration rate slows and follows the secular evolution. In this later evolution stage, the moon encounters fewer MMRs as the gap between MMRs widens at lower values of N:1. In addition to the eccentricity pumping, the moon’s semi-major axis $a_{\rm m}$ oscillates (see Fig.~\ref{fig:m2_time}b and Fig.~\ref{fig:m2_time}c) at later times, indicative of strong interactions with the MMRs or strong kicks directly from the host star.

A typical bump in the moon’s semi-major axis is $\sim0.035$ to $\sim0.050$ $R_{\rm H}$ due to the MMRs and other effects. We apply this bump to the secular tidal evolution to mimic these extra effects on top of our secular results, where for the M2 case with very weak tides ($\tau_{\rm p} = 10\ {\rm s}$, Fig. \ref{fig:m2_param}d), only a few parameters remain marginally stable. It is important to note that a $\tau_{\rm p}$ of 10 seconds may require an exotic (maybe nonphysical) interior for the planet considered here and is therefore unlikely. 

\begin{figure*}[ht!]
\plotone{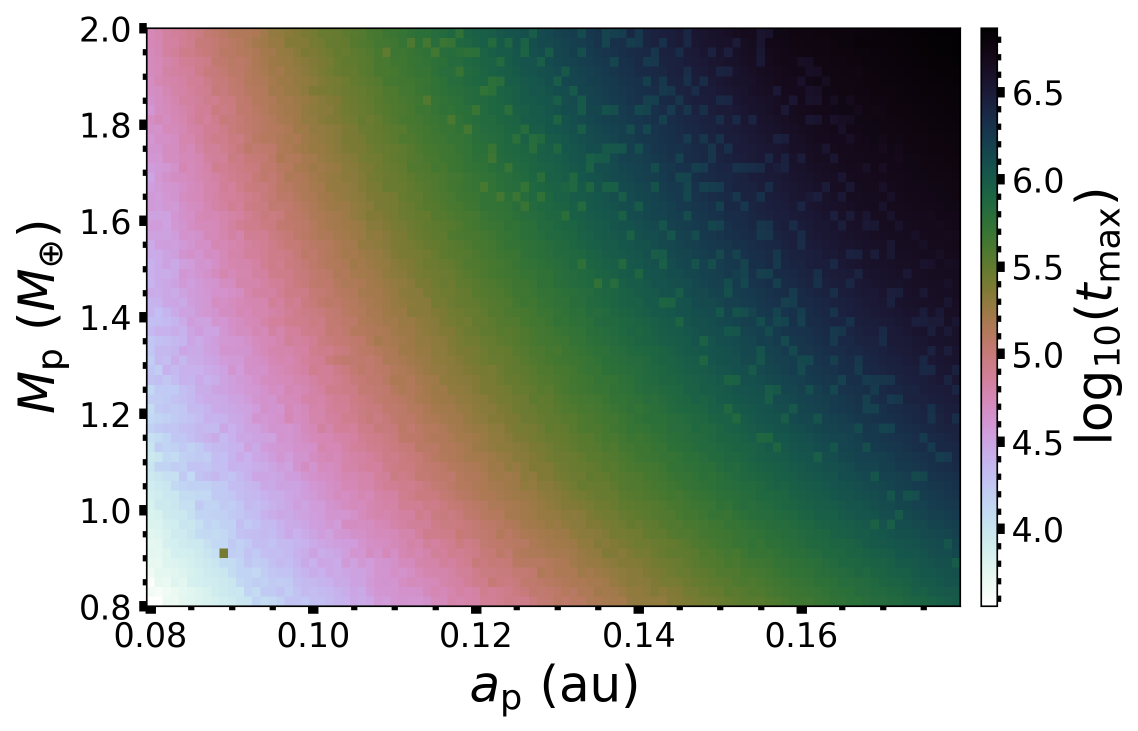}
\caption{Stability simulation results for a HZ planet--moon orbiting an M4-dwarf. Logarithm of maximum moon lifetime ($\log_{10}\left[t_{\rm max}\right]$) from \texttt{rebound} simulations for considering a range of host planet masses (0.8 to 2 $M_\oplus$) and semi-major axis within the HZ of an M4-dwarf.
\label{fig:m4_param}}
\end{figure*}

The moon’s outward migration is fueled by the angular momentum transfer, where the planet’s rotational period after ${\sim}10$ Myr increases only up to ${\sim}7-11$ hr (Fig.~\ref{fig:m2_spin}) in the \texttt{rebound} simulations.  Previous studies \citep{barnes02,sasaki12,piro18} using the constant $Q$ model for tides showed that moons can escape planets even if they start with with more slowly rotating planets. Our simulations use a stability limit prescription \citep[i.e.,][]{Rosario-Franco2020} that excludes the chaotic transition region that is phase-dependent, but if we consider larger stability limit up to ${\sim}0.5\ R_{\rm H}$ \citep[i.e.,][]{domingos06}, the outward migration will continue until this threshold is crossed ($\sim 100\ {\rm Myr})$. 

\subsection{Habitable Zone Moons in M4-dwarf systems}
For the case of an M4-dwarf,the HZ is closer-in and narrower, where we see a similar gradient in moon lifetimes with the planetary mass and initial semi-major axis. Figure ~\ref{fig:m4_param} shows our M4 results, but with a distinct cubehelix colormap (compared to the rainbow colormap in Fig. \ref{fig:m2_param}) to emphasize the differing colorbar limits.  Notably, all the simulations in this parameter space terminate before 10 million years (Fig.~\ref{fig:m4_param}). Since the HZ around an M4-dwarf is closer-in, the tidal effects from the host star are much stronger and the moon's migration is faster, allowing it to escape on a shorter timescale.  Additionally, we expect potential moons orbiting Earth-like planets within the HZs of later (M4+) dwarfs to showcase a similar behavior \citep{kane17}. By contrast, the HZs of earlier M-dwarfs the HZ extends farther out, which weakens tidal effects from the host star and opens the possibility of longer-lived moons.

\subsection{Habitable Zone Moons in M0-dwarf systems}
\begin{figure*}[ht!]
\plotone{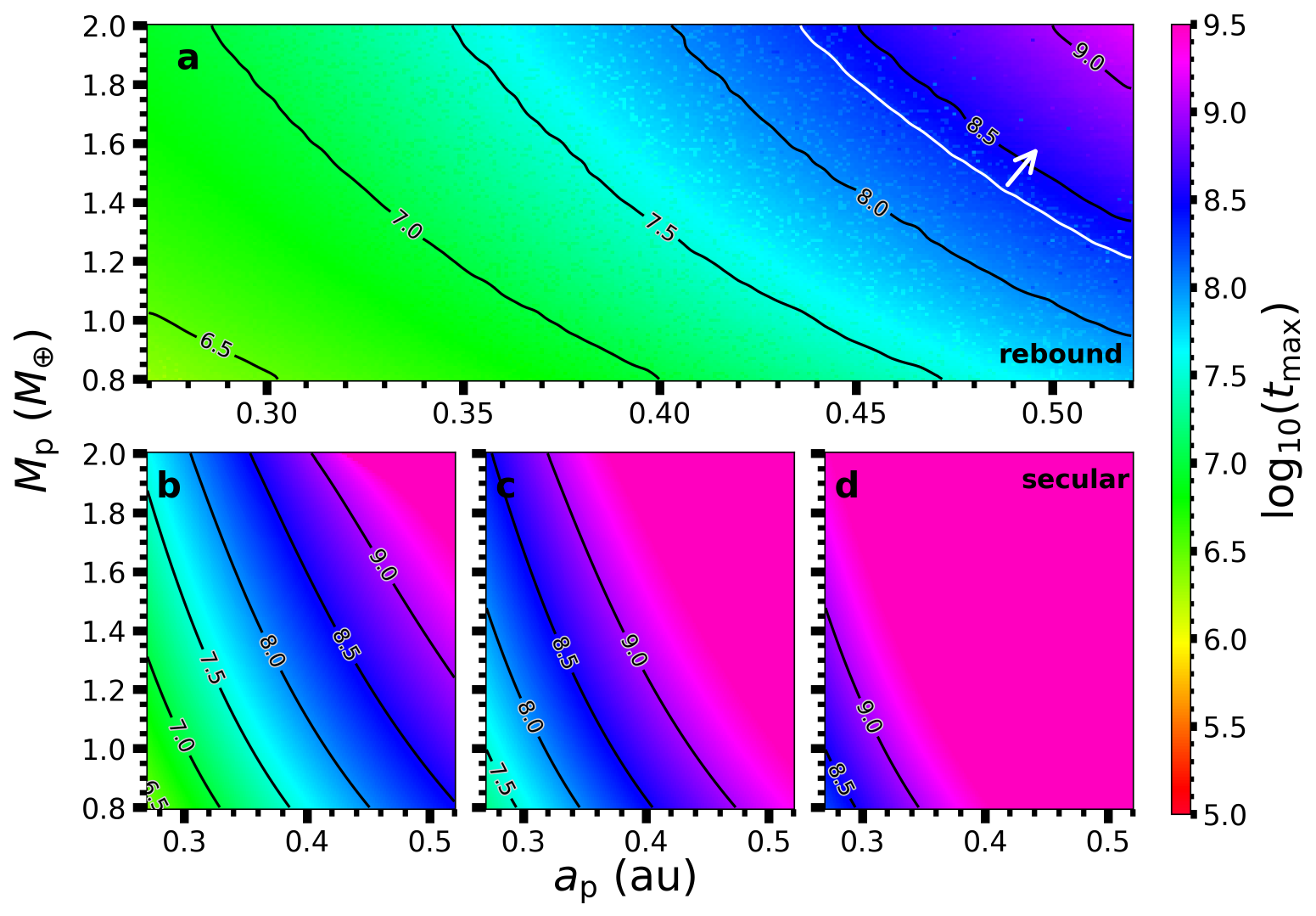}
\caption{Stability simulation results for a HZ planet--moon orbiting an M0-dwarf. Similar to Fig.~\ref{fig:m2_param} but for moons orbiting an Earth-sized planet in an M0 system. The white contour in (a) marks the boundary beyond which the lifetimes are extrapolated, as indicated by the arrow. The maximum lifetime in (a) is $\approx$ 9.2 in $\log_{10}(t)$.
\label{fig:m0_param}}
\end{figure*}

For an M0-dwarf, the HZ extends farther out and spans a wider range because of the star’s higher luminosity. This geometry shifts the balance, where moons can survive much longer before reaching instability. As a result, there is a shift towards longer lifetimes due to the longer orbital periods and the overall weaker stellar tide at larger distances (Fig.~\ref{fig:m0_param}). We model systems with Earth-like tides ($\tau_{\rm p} = 698\ {\rm s}$) up to 200 Myr to capture the outward migration from ${\sim}$ $8-60$\ $R_{\rm p}$.  The stability limit for the outer HZ in these systems is ${\sim} 66$\ $R_{\rm p}$ \citep{Rosario-Franco2020}, where those simulations that survive for 200 Myr are shown by the upper right region beyond the white contour (Fig.~\ref{fig:m0_param}a).

\begin{figure*}[ht!]
\plotone{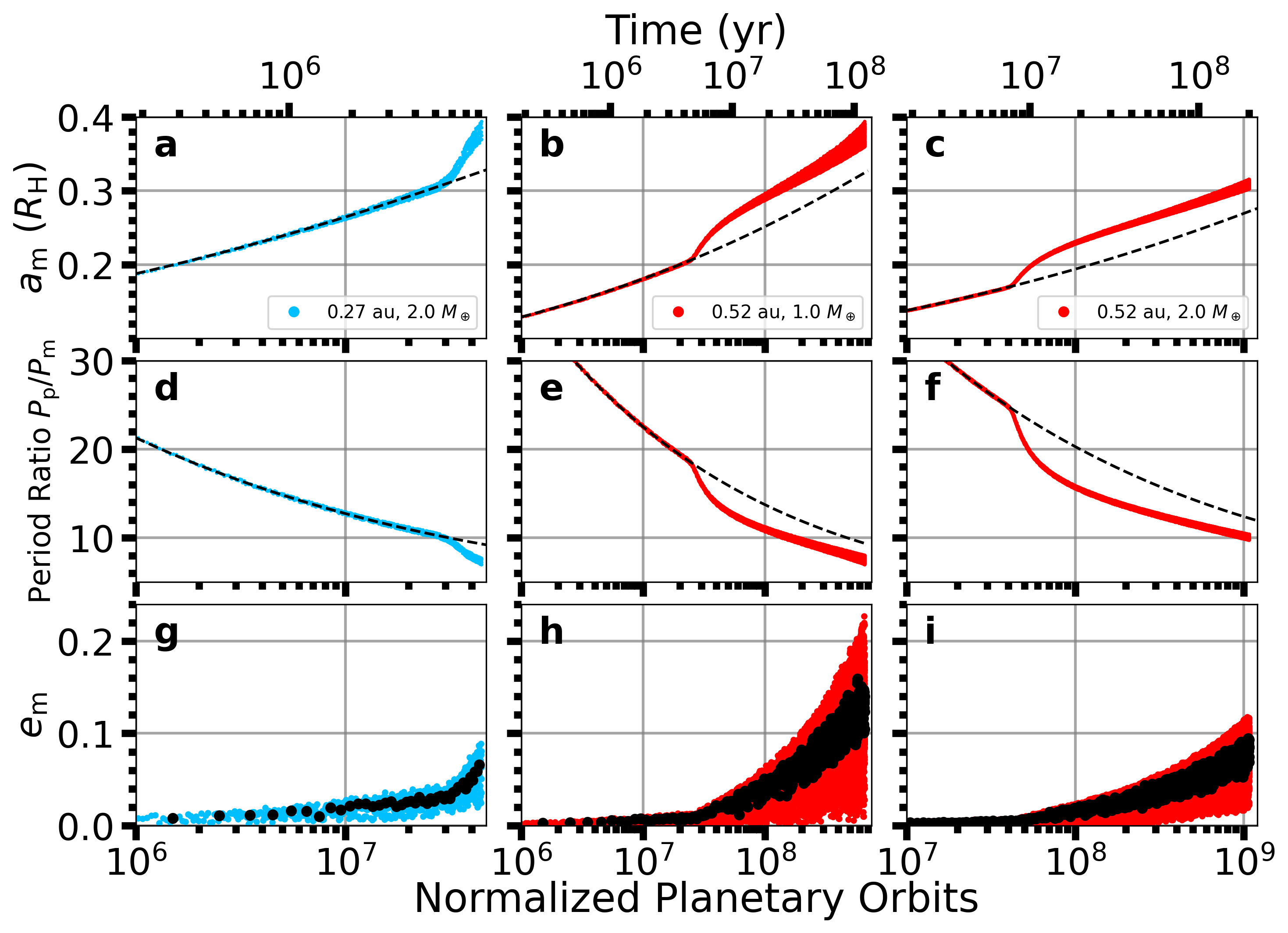}
\caption{
Time series plot for three HZ planet--moon systems orbiting an M0-dwarf. Same as Fig.~\ref{fig:m2_time} but for moons in an M0 system. The 3 distinct points used are (0.27 au and 2.0 $M_\oplus$, 0.52 au and 1.0 $M_\oplus$, 0.52 au and 2.0 $M_\oplus$). The bottom $x$-axis uses 0.27 au as the inner edge of the HZ for normalization.
\label{fig:m0_time}}
\end{figure*}

Beyond the white contour, we use polynomial fits to extrapolate and estimate the lifetimes of these cases. Similar to our results for an M2-dwarf (see \S \ref{sec:m2_systems}, a bump in migration rate due to MMRs appears just before ${\sim}10\ {\rm Myr}$ (see Fig.~\ref{fig:m0_time}). We therefore apply a best-fit line (using \texttt{numpy.polyfit}; \citet{harris2020}) to the later stages of the moon's evolution, when tidal migration proceeds quasi-linearly in logarithmic time. To reduce bias from the starting point, we perform 1,000 fits with intervals beginning at ${\sim}20-100$ Myr and extending to 200 Myr.

\begin{figure*}[ht!]
\includegraphics[scale=0.8]{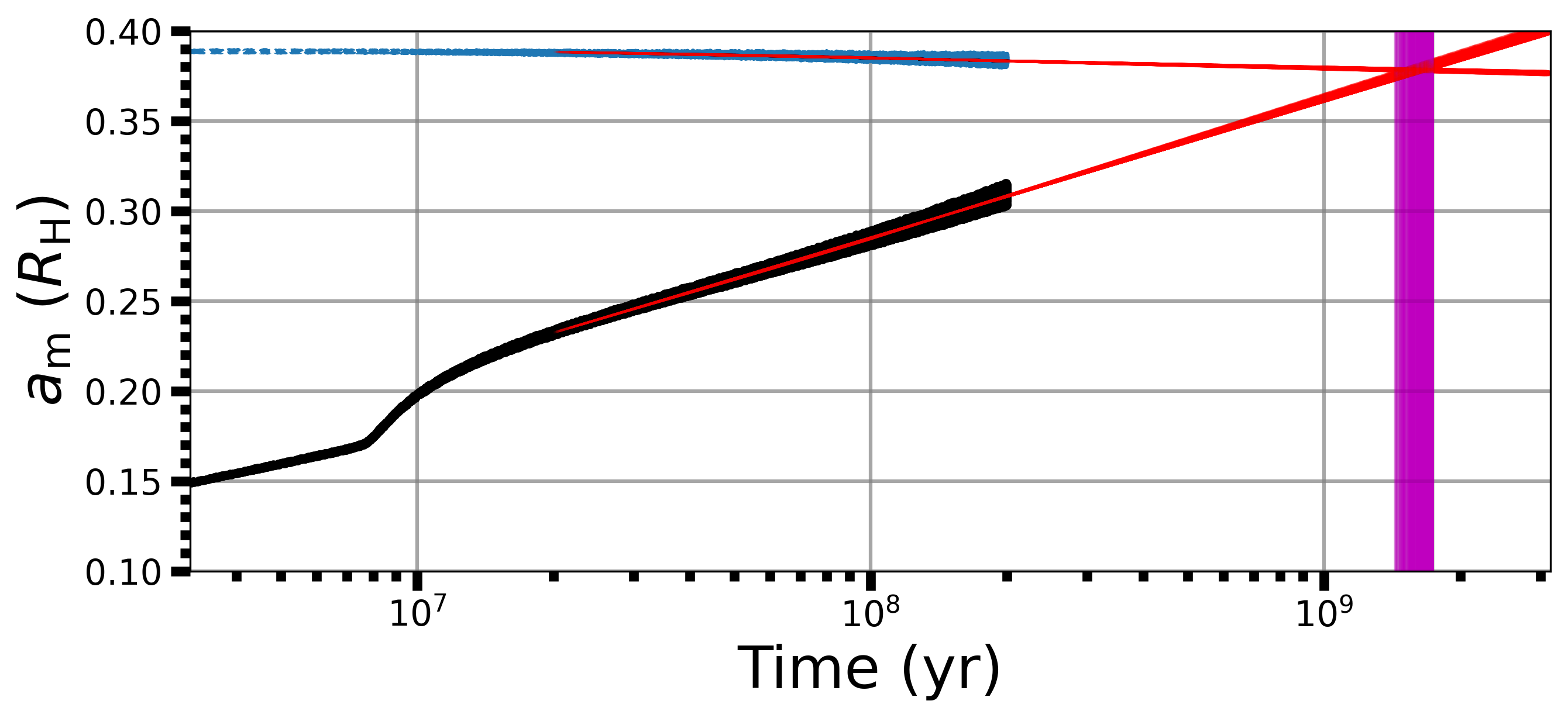}
\caption{Best-fit analysis for the lifetime of one HZ planet--moon system orbiting an M0-dwarf. Time evolution of the exomoon’s semi-major axis for a representative case within the M0 parameter space (0.52 au, 2.0 $M_\oplus$), computed using \texttt{rebound}. Black points denote the \texttt{rebound} simulation results, while the blue dashed line indicates the moon’s stability limit. Red curves correspond to best-fit lines for both the stability threshold and the \texttt{rebound} evolution. Their intersection, marking the estimated instability time, is highlighted in magenta.
\label{fig:m0_bestfit}}
\end{figure*}

In addition, we account for changes in the moon’s eccentricity over the same interval and its effect on the computed stability limit \citep{Rosario-Franco2020}, incorporating the effects of both moon and planetary eccentricities ($e_{\rm m}$ and $e_{\rm p}$). The intersection point of the best-fit line for the moon’s extrapolated semi-major axis and the stability limit estimates the time when the moon reaches the stability threshold (Fig.~\ref{fig:m0_bestfit}). Each of the 1,000 fits has a slightly different slope, which results in a range in the projected lifetime.  The longest-lived system lies at the upper right of the $M_{\rm p} - a_{\rm p}$ parameter space (i.e., 0.52 AU, 2 $M_\oplus$), where we find a mean lifetime of $\sim$1.61 Gyr with a 1$\sigma$ confidence interval of roughly $1.52 – 1.70$ Gyr.

When second-order polynomial fits are applied, the estimated mean lifetime decreases to $\approx$1.16 Gyr, with a corresponding 1$\sigma$ range of $\approx$1.05–1.26 Gyr. Higher-order polynomial fits are possible but lack physical plausibility because they can have a negative curvature. We therefore adopt the first-order fits, which provide the most optimistic lifetime and physically interpretable results. To assess the robustness of these best-fit calculations, we repeat the analysis for the region below the white curve, focusing specifically on cases with lifetimes in the 30–100 Myr range and compare the estimated lifetimes to actual values obtained from simulations.

On average, the best-fit approach yields a lifetime estimate that is approximately 16\% longer than the simulated values, with a 1$\sigma$ range of 9.3\% to 23.6\%. As the eccentricity grows (see Figs.~\ref{fig:m0_time}g-i), the semi-major axis evolution can accelerate toward the end of the moon’s lifetime, resulting in a migration rate faster than the simple best-fit model. If we correct for this bias in our original results, the revised longest-lived estimate is $\approx$1.35 Gyr.

We further probe the outward migration of the moon by exploring the time series evolution for a few points in the M0 parameter space in a similar manner as in Sec. \ref{sec:m2_systems}. If we consider a moon-hosting planet at the inner edge of the HZ (0.27 au, 2.0 $M_\oplus$; Fig.~\ref{fig:m0_time}a), we see a relatively smooth evolution of $a_{\rm m}$ until around 5 Myr where the planet--moon period ratio reaches about 10:1 and MMRs induce a bump in the migration rate (Fig.~\ref{fig:m0_time}d).
If we instead consider Earth-like planets ($1\ M_\oplus$ or $2\ M_\oplus$) orbiting at the outer edge of the HZ ($0.52\ {\rm au}$, the evolution of $a_{\rm m}$ (Figs.~\ref{fig:m0_time}b and \ref{fig:m0_time}c) also has a noticeable bump due to the exposure to MMRs; however, the boost in migration is not enough to bring the moon to the stability limit in the time allowed for the more massive ($2\ M_\oplus$) case. We note that this bump is also on the order of 0.035 $R_{\rm H}$ where the evolution of the moon then stabilizes to a similar slope to the secular theory afterwards.

From a secular tide model assuming Earth-like tides, we find the vast majority of the initial conditions are unstable within 5 Gyr, barring the top right portion (Fig.~\ref{fig:m0_param}b). When adding the bump of 0.035 $R_{\rm H}$, we find these survivors should migrate past the stability limit to eventually become unstable. In the weak tide limit ($\tau_{\rm p} = 100\ {\rm s}$; Fig.~\ref{fig:m0_param}c), the top right region with billion-year lifetimes grows larger and after adding the bump, we find that only a small region of the $M_{\rm p}-a_{\rm p}$ parameter space remains within the stability limit. For the very weak tide case ($\tau_{\rm p} = 10\ {\rm s}$; Fig.~\ref{fig:m0_param}d), the majority of lifetimes are billions of years.  Such a small tidal lag implies an unusual planetary interior, requiring further study to assess plausibility.

\begin{figure*}[ht!]
\plotone{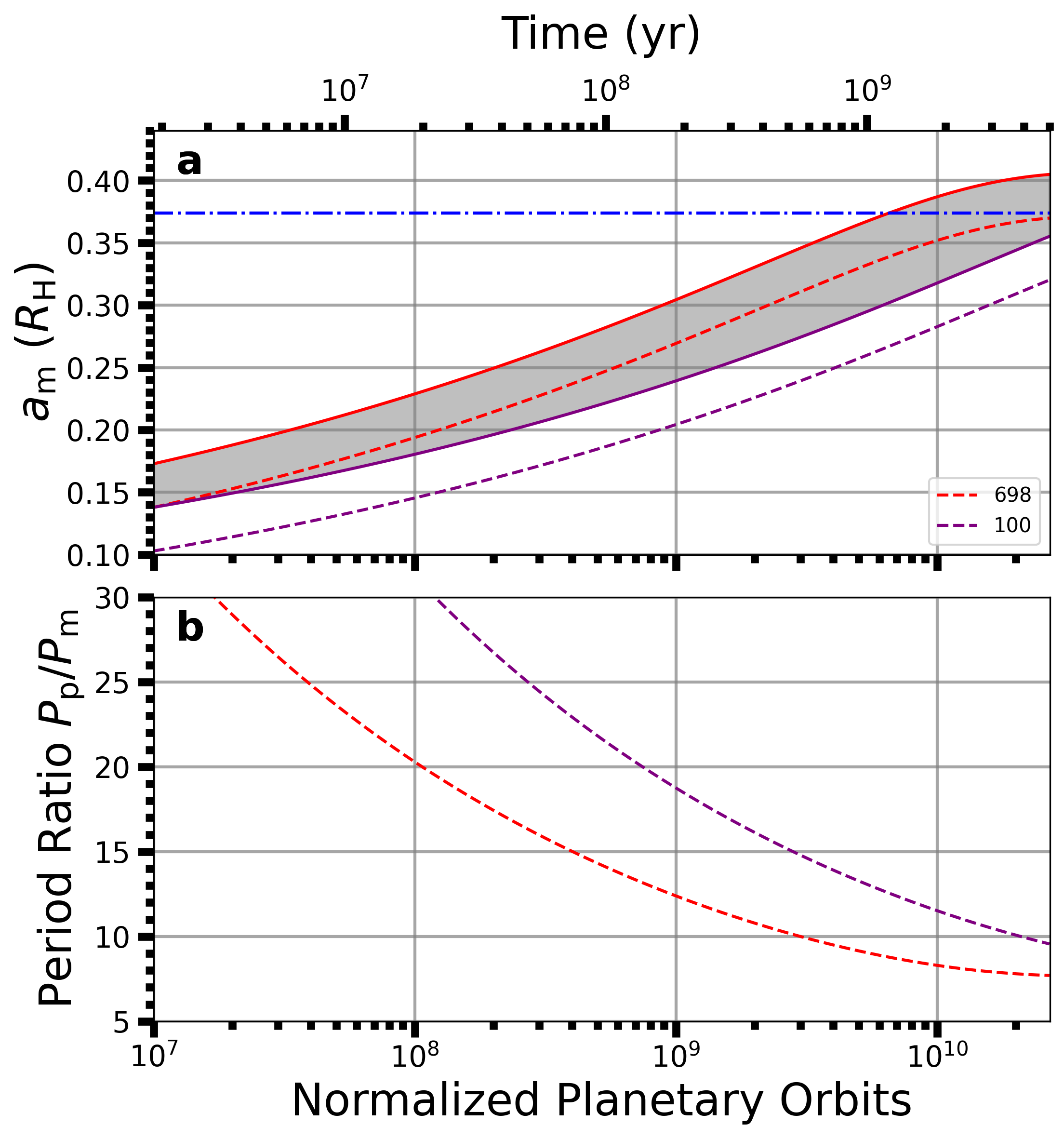}
\caption{Secular evolution of one HZ planet--moon system orbiting an M0-dwarf. Time evolution of the (a) exomoon's semi-major axis $a_{\rm m}$ and (b) planet to moon period ratio $P_{\rm p}/P_{\rm m}$ for one case in the M0 parameter space (0.52 au, 2.0 $M_\oplus$) using secular tidal theory. The bottom $x$-axis, plotted on a logarithmic scale, shows time normalized by the orbital period of a planet at the inner edge of the HZ (0.27 au) while the top $x$-axis shows time (in years). The red and purple dashed lines represent a $\tau_{\rm p}$ of 698 and 100 seconds, respectively, while the red and purple solid lines represent those curves with the added bump of $0.035\ R_{\rm H}$, accounting for expected MMR and 3-body effects. The gray shaded region represents the area of expected curves between the boosted 698 second upper limit and boosted 100 second lower limit while the blue dash-dot line represents the moon's stability limit.
\label{fig:m0_sec}}
\end{figure*}

We examine the secular tidal migration considering a 2.0 $M_\oplus$ planet orbiting at the outer edge of the HZ (0.52 au) because it has the slowest evolution (Fig. \ref{fig:m0_sec}), where it takes $1\ {\rm Gyr}$ for the moon to reach $\sim$ 0.33 $R_{\rm H}$. By incorporating the bump in the migration rate into our secular results, the moon will migrate to $\sim$ $0.365\ R_{\rm H}$ within 1 Gyr, which closely approaches the stability threshold (Fig.~\ref{fig:m0_sec}a) and crosses the stability limit by 1.3 Gyr. This lifetime estimate is comparable to the estimate from our extrapolation method, further reinforcing the system's susceptibility to instability. 

\section{Summary and Conclusions} \label{sec:summary}

Our findings suggest that HZ Earth-like planets in M-dwarf systems will lose large (Luna-like) moon(s) (if formed) within the first billion years of their existence.  The planetary HZs vary depending on the spectral subtype of the host star, which in turn varies the strength of the stellar tides.  Our 200 Myr numerical simulations of star--planet--moon systems orbiting the HZ of M4-dwarfs show the typical moon lifetime is less than 10 Myr (Fig.~\ref{fig:m4_param}), which is very short compared to the astrobiological, geological, or astrophysical timescales.  Given these outcomes, we expect similar systems orbiting in the HZs of later M-dwarfs (M5-M9) to lose their moon(s) on even shorter timescales \citep{kane17}.

These results echo earlier work on giant-planet moons in M-dwarf systems. Using tidal and dynamical models, \citet{zoll17} showed that massive exomoons are unlikely to sustain habitable conditions due to extreme tidal heating, and that moons orbiting lower-mass M-dwarfs are dynamically unstable. Together with our findings, this points to a general fragility of exomoons in M-dwarf systems. An early M2-dwarf system does not provide a haven for moon(s) either, where such a moon can survive a while longer.  The moon’s outward migration will cross the stability limit \citep{Rosario-Franco2020} within 1--50 Myr. The broad range of moon lifetimes depends on the planet’s placement within the HZ and the assumed composition through the tidal time-lag.

A large moon’s lifetime can reach 1 Gyr if it orbits a habitable Earth-mass planet orbiting an M0-dwarf.  This configuration is special because its outer HZ edge extends to 0.52 au, which weakens the stellar tide on the host planet and leaves a large part of the tidal action to the moon’s tide that despins the host planet.  When considering an Earth-like tidal time-lag ($\tau_{\rm p} = 698$~s; possibly an ocean-world Earth) and a more massive host planet ($M_{\rm p}=2\ M_\oplus$), we find a maximum moon lifetime of 1.35 Gyr.  This timescale corresponds to a time on the Earth when oxygen was just beginning to buildup in our atmosphere.

Even longer lifetimes ($>$5 Gyr) may be possible, but these require lower-mass moons or an exotic host planet composition so that the tide is less effective ($\tau_{\rm p}\lesssim 100$; a dry-Earth or weak tide).  Higher-mass planets located in the outer regions of the HZs within M0-type stellar systems may be the only configuration to host large, observable (and potentially habitable \citep{marrod19}) moons. Smaller, low-mass moons (e.g., Phobos / Ceres-like) would still be possible; however, if present, they would exist beyond the current limits of photometric detectability (relative to the noise floor) or reveal themselves through transit-timing variations \citep{Kipping2009}.  If a moon is indeed required for complex life to develop on a host planet, our results suggest that this limitation may represent an additional contributing factor to the Drake equation and Fermi paradox, as M-dwarfs dominate the stellar population yet appear largely unable to host large, HZ moons.

It is important to note that M0–M4 dwarfs have pre-main-sequence lifetimes of $\approx$ 200–400 Myr \citep{ram14}. Throughout this phase, planets orbiting these stars are subjected to intense flaring activity and elevated UV flux, conditions that may inhibit or delay the emergence of life \citep{luger15, getman22}. Under such circumstances, large moons orbiting these planets would likely already have become dynamically unstable prior to the emergence of life in this delayed timeline.

Additionally, when considering planetary occurrence rates around M-dwarfs, we expect on average one to two Earth-sized planets per star, with a bias toward short-period systems \citep{Gaidos16, Hard19}. For HZ planets, \citet{dressing15} estimate an occurrence rate of 0.24 Earth-sized planets per M-dwarf, while \citet{hsu20} report a rate of 0.33 for planets with radii between 0.75 and 1.0 $R_{\oplus}$. As future surveys shift focus toward longer-period HZ planets, Earth-like planets orbiting within the HZs of M-dwarfs are expected to be a central part of future observational campaigns. 

\section{Outlook} \label{sec:outlook}
The Habitable Worlds Observatory (HWO), a proposed $6-8$ meter space telescope, is a promising tool in the search for exomoons, in addition to its main task of finding habitable Earth-like exoplanets \citep{HWO}. With this telescope the detection of Luna-sized exomoons may be possible \citep{limbach24}. Given our results, such a telescope might (1) detect HZ moons orbiting Earth-like planets around M0-dwarfs (depending on the system's age) and (2) provide observational verification (according to the instrumental detection limit) for the absence of HZ exomoons around those planets in systems with host stars of spectral type M2 or later.   This mission is built on the concepts of two other proposed space telescopes, LUVOIR \citep{luvo19} and HabEx \citep{gaud20}. 

Moreover, the Giant Magellan Telescope (GMT), a $25.4$ meter ground-based observatory currently under construction will offer improved spatial resolution and sensitivity, allowing for the direct imaging of exoplanets and the possible detection of exomoons through methods such as direct imaging \citep{GMT}.  In addition to established methods used in exoplanet detection such as photometry and transit timings \citep{teachey2024}, emerging observational techniques may also prove useful in the detection of exomoons. \citet{vanderburg18} provides novel radial velocity (RV) methods combined with direct imaging while cyclotron radio emission methods \citet{noyola14,noyola16} could emerge useful in exomoon observations.

Two programs were selected for observation in JWST cycle 3 including a search around: (1) a long-period Jupiter-like planet orbiting a K-dwarf \citep[GO 6491]{Cassese2024} and (2) a terrestrial HZ planet orbiting an M-dwarf \citep[GO 6193]{jwst24}, which provide two different avenues of exploration.  Our work suggests that program (2) would not be successful given the estimated age of TOI-700 is $>1.5\ {\rm Gyr}$, the host star is a spectral type M2V, and the host planet (TOI-700d) lies at 0.16 au \citep{Gilbert2020}.  Even within the very weak tide scenario, the expected moon lifetime is $\lesssim 100\ {\rm Myr}$ (see Fig. \ref{fig:m2_param}).  Program (1) will attempt to detect Galilean sized moons, which has its own challenges \citep[see][]{Millholland2025}, but escape via tides is of no concern.

\begin{acknowledgments}
    The authors thank the anonymous reviewer whose comments greatly improved the quality of the manuscript. S.D.P. and N.N.W. acknowledge support by the National Science Foundation (NSF) under grant No. AST-2054353.  B.Q. acknowledges support in part by the Texas A\&M High Performance Research Computing (HPRC) and the NSF under grant No. 2232895. The authors acknowledge the Texas A\&M HPRC for providing computing resources on the Launch cluster that contributed to the research results reported here.
\end{acknowledgments}

\software{\texttt{rebound} \citep{rebound}, \texttt{reboundx} \citep{reboundx}}

\bibliography{tides_arxiv}{}

\begin{thebibliography}{}
\expandafter\ifx\csname natexlab\endcsname\relax\def\natexlab#1{#1}\fi
\providecommand{\url}[1]{\href{#1}{#1}}
\providecommand{\dodoi}[1]{doi:~\href{http://doi.org/#1}{\nolinkurl{#1}}}
\providecommand{\doeprint}[1]{\href{http://ascl.net/#1}{\nolinkurl{http://ascl.net/#1}}}
\providecommand{\doarXiv}[1]{\href{https://arxiv.org/abs/#1}{\nolinkurl{https://arxiv.org/abs/#1}}}

\bibitem[{J.~W. Barnes \& D.~P. O’Brien(2002)Barnes \& O’Brien}]{barnes02}
Barnes, J.~W., \& O’Brien, D.~P. 2002, \bibinfo{title}{Stability of satellites around close-in extrasolar giant planets,} \apj, 575, 1087, \dodoi{10.1086/341477}

\bibitem[{R. {Barnes}(2017){Barnes}}]{Barnes_2017}
{Barnes}, R. 2017, \bibinfo{title}{{Tidal locking of habitable exoplanets},} Celestial Mechanics and Dynamical Astronomy, 129, 509, \dodoi{10.1007/s10569-017-9783-7}

\bibitem[{R.~A. {Bernstein} {et~al.}(2014){Bernstein}, {McCarthy}, {Raybould}, {Bigelow}, {Bouchez}, {Filgueira}, {Jacoby}, {Johns}, {Sawyer}, {Shectman}, \& {Sheehan}}]{GMT}
{Bernstein}, R.~A., {McCarthy}, P.~J., {Raybould}, K., {et~al.} 2014, \bibinfo{title}{{Overview and status of the Giant Magellan Telescope project},} in Society of Photo-Optical Instrumentation Engineers (SPIE) Conference Series, Vol. 9145, Ground-based and Airborne Telescopes V, ed. L.~M. {Stepp}, R.~{Gilmozzi}, \& H.~J. {Hall}, 91451C, \dodoi{10.1117/12.2055282}

\bibitem[{ {Bolmont, Emeline} {et~al.}(2015){Bolmont, Emeline}, {Raymond, Sean N.}, {Leconte, Jeremy}, {Hersant, Franck}, \& {Correia, Alexandre C. M.}}]{bolmont15}
{Bolmont, Emeline}, {Raymond, Sean N.}, {Leconte, Jeremy}, {Hersant, Franck}, \& {Correia, Alexandre C. M.} 2015, \bibinfo{title}{Mercury-T: A new code to study tidally evolving multi-planet systems. Applications to Kepler-62⋆,} \aap, 583, A116, \dodoi{10.1051/0004-6361/201525909}

\bibitem[{M.~J. {Burchell}(2006){Burchell}}]{burchell06}
{Burchell}, M.~J. 2006, \bibinfo{title}{{W(h)ither the Drake equation?},} International Journal of Astrobiology, 5, 243, \dodoi{10.1017/S1473550406003107}

\bibitem[{R.~M. {Canup}(2004){Canup}}]{Canup2004}
{Canup}, R.~M. 2004, \bibinfo{title}{{Dynamics of Lunar Formation},} \araa, 42, 441, \dodoi{10.1146/annurev.astro.41.082201.113457}

\bibitem[{B. {Cassese} {et~al.}(2024){Cassese}, {Batygin}, {Chachan}, {Changeat}, {Constantinou}, {Edwards}, {Kipping}, {Madhusudhan}, {Poddar}, {Teachey}, {Tinetti}, \& {Vega}}]{Cassese2024}
{Cassese}, B., {Batygin}, K., {Chachan}, Y., {et~al.} 2024, {Revealing the Oblateness and Satellite System of an Extrasolar Jupiter Analog},, JWST Proposal. Cycle 3, ID. \#6491

\bibitem[{G. {Chabrier}(2003){Chabrier}}]{chab03}
{Chabrier}, G. 2003, \bibinfo{title}{{Galactic stellar and substellar initial mass function},} \pasp, 115, 763, \dodoi{10.1086/376392}

\bibitem[{M. {Civiletti}(2025){Civiletti}}]{civiletti25}
{Civiletti}, M. 2025, \bibinfo{title}{{Quantifying the Fermi paradox via passive SETI: a general framework},} arXiv e-prints, arXiv:2505.00062, \dodoi{10.48550/arXiv.2505.00062}

\bibitem[{M. {Cuntz} {et~al.}(2013){Cuntz}, {Quarles}, {Eberle}, \& {Shukayr}}]{cuntz13}
{Cuntz}, M., {Quarles}, B., {Eberle}, J., \& {Shukayr}, A. 2013, \bibinfo{title}{{On the possibility of habitable moons in the system of HD 23079: Results from orbital stability studies},} \pasa, 30, e033, \dodoi{10.1017/pas.2013.011}

\bibitem[{R.~C. {Domingos} {et~al.}(2006){Domingos}, {Winter}, \& {Yokoyama}}]{domingos06}
{Domingos}, R.~C., {Winter}, O.~C., \& {Yokoyama}, T. 2006, \bibinfo{title}{{Stable satellites around extrasolar giant planets},} \mnras, 373, 1227, \dodoi{10.1111/j.1365-2966.2006.11104.x}

\bibitem[{F. Drake \& D. Sobel(1993)Drake \& Sobel}]{drake93}
Drake, F., \& Sobel, D. 1993, Is Anyone Out There?: The Scientific Search for Extraterrestrial Intelligence (Souvenir P.).
\newblock \url{https://books.google.com/books?id=ZrddHY0mQpgC}

\bibitem[{C.~D. {Dressing} \& D. {Charbonneau}(2015){Dressing} \& {Charbonneau}}]{dressing15}
{Dressing}, C.~D., \& {Charbonneau}, D. 2015, \bibinfo{title}{{The occurrence of potentially habitable planets orbiting M dwarfs estimated from the full Kepler dataset and an empirical measurement of the detection sensitivity},} \apj, 807, 45, \dodoi{10.1088/0004-637X/807/1/45}

\bibitem[{P.~P. {Eggleton} {et~al.}(1998){Eggleton}, {Kiseleva}, \& {Hut}}]{eggleton}
{Eggleton}, P.~P., {Kiseleva}, L.~G., \& {Hut}, P. 1998, \bibinfo{title}{{The equilibrium tide model for tidal friction},} \apj, 499, 853, \dodoi{10.1086/305670}

\bibitem[{E. {Gaidos} {et~al.}(2016){Gaidos}, {Mann}, {Kraus}, \& {Ireland}}]{Gaidos16}
{Gaidos}, E., {Mann}, A.~W., {Kraus}, A.~L., \& {Ireland}, M. 2016, \bibinfo{title}{{They are small worlds after all: revised properties of Kepler M dwarf stars and their planets},} \mnras, 457, 2877, \dodoi{10.1093/mnras/stw097}

\bibitem[{B.~S. Gaudi {et~al.}(2020)Gaudi, Seager, Mennesson, Kiessling, Warfield, Cahoy, Clarke, Domagal-Goldman, Feinberg, Guyon, Kasdin, Mawet, Plavchan, Robinson, Rogers, Scowen, Somerville, Stapelfeldt, Stark, Stern, Turnbull, Amini, Kuan, Martin, Morgan, Redding, Stahl, Webb, Alvarez-Salazar, Arnold, Arya, Balasubramanian, Baysinger, Bell, Below, Benson, Blais, Booth, Bourgeois, Bradford, Brewer, Brooks, Cady, Caldwell, Calvet, Carr, Chan, Cormarkovic, Coste, Cox, Danner, Davis, Dewell, Dorsett, Dunn, East, Effinger, Eng, Freebury, Garcia, Gaskin, Greene, Hennessy, Hilgemann, Hood, Holota, Howe, Huang, Hull, Hunt, Hurd, Johnson, Kissil, Knight, Kolenz, Kraus, Krist, Li, Lisman, Mandic, Mann, Marchen, Marrese-Reading, McCready, McGown, Missun, Miyaguchi, Moore, Nemati, Nikzad, Nissen, Novicki, Perrine, Pineda, Polanco, Putnam, Qureshi, Richards, Riggs, Rodgers, Rud, Saini, Scalisi, Scharf, Schulz, Serabyn, Sigrist, Sikkia, Singleton, Shaklan, Smith, Southerd, Stahl, Steeves, Sturges, Sullivan, Tang,
  Taras, Tesch, Therrell, Tseng, Valente, Buren, Villalvazo, Warwick, Webb, Westerhoff, Wofford, Wu, Woo, Wood, Ziemer, Arney, Anderson, Maíz-Apellániz, Bartlett, Belikov, Bendek, Cenko, Douglas, Dulz, Evans, Faramaz, Feng, Ferguson, Follette, Ford, García, Geha, Gelino, Götberg, Hildebrandt, Hu, Jahnke, Kennedy, Kreidberg, Isella, Lopez, Marchis, Macri, Marley, Matzko, Mazoyer, McCandliss, Meshkat, Mordasini, Morris, Nielsen, Newman, Petigura, Postman, Reines, Roberge, Roederer, Ruane, Schwieterman, Sirbu, Spalding, Teplitz, Tumlinson, Turner, Werk, Wofford, Wyatt, Young, \& Zellem}]{gaud20}
Gaudi, B.~S., Seager, S., Mennesson, B., {et~al.} 2020, The Habitable Exoplanet Observatory (HabEx) Mission concept study final report, \doarXiv{2001.06683}

\bibitem[{K.~V. {Getman} {et~al.}(2022){Getman}, {Feigelson}, {Garmire}, {Broos}, {Kuhn}, {Preibisch}, \& {Airapetian}}]{getman22}
{Getman}, K.~V., {Feigelson}, E.~D., {Garmire}, G.~P., {et~al.} 2022, \bibinfo{title}{{Evolution of x-ray activity in <25 Myr old pre-main sequence stars},} \apj, 935, 43, \dodoi{10.3847/1538-4357/ac7c69}

\bibitem[{E.~A. {Gilbert} {et~al.}(2020){Gilbert}, {Barclay}, {Schlieder}, {Quintana}, {Hord}, {Kostov}, {Lopez}, {Rowe}, {Hoffman}, {Walkowicz}, {Silverstein}, {Rodriguez}, {Vanderburg}, {Suissa}, {Airapetian}, {Clement}, {Raymond}, {Mann}, {Kruse}, {Lissauer}, {Col{\'o}n}, {Kopparapu}, {Kreidberg}, {Zieba}, {Collins}, {Quinn}, {Howell}, {Ziegler}, {Vrijmoet}, {Adams}, {Arney}, {Boyd}, {Brande}, {Burke}, {Cacciapuoti}, {Chance}, {Christiansen}, {Covone}, {Daylan}, {Dineen}, {Dressing}, {Essack}, {Fauchez}, {Galgano}, {Howe}, {Kaltenegger}, {Kane}, {Lam}, {Lee}, {Lewis}, {Logsdon}, {Mandell}, {Monsue}, {Mullally}, {Mullally}, {Paudel}, {Pidhorodetska}, {Plavchan}, {Reyes}, {Rinehart}, {Rojas-Ayala}, {Smith}, {Stassun}, {Tenenbaum}, {Vega}, {Villanueva}, {Wolf}, {Youngblood}, {Ricker}, {Vanderspek}, {Latham}, {Seager}, {Winn}, {Jenkins}, {Bakos}, {Brice{\~n}o}, {Ciardi}, {Cloutier}, {Conti}, {Couperus}, {Di Sora}, {Eisner}, {Everett}, {Gan}, {Hartman}, {Henry}, {Isopi}, {Jao}, {Jensen}, {Law}, {Mallia},
  {Matson}, {Shappee}, {Le Wood}, \& {Winters}}]{Gilbert2020}
{Gilbert}, E.~A., {Barclay}, T., {Schlieder}, J.~E., {et~al.} 2020, \bibinfo{title}{{The First Habitable-zone Earth-sized Planet from TESS. I. Validation of the TOI-700 System},} \aj, 160, 116, \dodoi{10.3847/1538-3881/aba4b2}

\bibitem[{E.~A. {Gilbert} {et~al.}(2023){Gilbert}, {Vanderburg}, {Rodriguez}, {Hord}, {Clement}, {Barclay}, {Quintana}, {Schlieder}, {Kane}, {Jenkins}, {Twicken}, {Kunimoto}, {Vanderspek}, {Arney}, {Charbonneau}, {G{\"u}nther}, {Huang}, {Isopi}, {Kostov}, {Kristiansen}, {Latham}, {Mallia}, {Mamajek}, {Mireles}, {Quinn}, {Ricker}, {Schulte}, {Seager}, {Suissa}, {Winn}, {Youngblood}, \& {Zapparata}}]{Gilbert2023}
{Gilbert}, E.~A., {Vanderburg}, A., {Rodriguez}, J.~E., {et~al.} 2023, \bibinfo{title}{{A Second Earth-sized Planet in the Habitable Zone of the M Dwarf, TOI-700},} \apjl, 944, L35, \dodoi{10.3847/2041-8213/acb599}

\bibitem[{K.~K. {Hardegree-Ullman} {et~al.}(2019){Hardegree-Ullman}, {Cushing}, {Muirhead}, \& {Christiansen}}]{Hard19}
{Hardegree-Ullman}, K.~K., {Cushing}, M.~C., {Muirhead}, P.~S., \& {Christiansen}, J.~L. 2019, \bibinfo{title}{{Kepler planet occurrence rates for mid-type M dwarfs as a function of spectral type},} \aj, 158, 75, \dodoi{10.3847/1538-3881/ab21d2}

\bibitem[{C.~R. Harris {et~al.}(2020)Harris, Millman, van~der Walt, Gommers, Virtanen, Cournapeau, Wieser, Taylor, Berg, Smith, Kern, Picus, Hoyer, van Kerkwijk, Brett, Haldane, del R{\'{i}}o, Wiebe, Peterson, G{\'{e}}rard-Marchant, Sheppard, Reddy, Weckesser, Abbasi, Gohlke, \& Oliphant}]{harris2020}
Harris, C.~R., Millman, K.~J., van~der Walt, S.~J., {et~al.} 2020, \bibinfo{title}{Array programming with {NumPy},} Nature, 585, 357, \dodoi{10.1038/s41586-020-2649-2}

\bibitem[{R. {Heller} {et~al.}(2011){Heller}, {Leconte}, \& {Barnes}}]{heller11}
{Heller}, R., {Leconte}, J., \& {Barnes}, R. 2011, \bibinfo{title}{{Tidal obliquity evolution of potentially habitable planets},} \aap, 528, A27, \dodoi{10.1051/0004-6361/201015809}

\bibitem[{R. {Heller} {et~al.}(2014){Heller}, {Williams}, {Kipping}, {Limbach}, {Turner}, {Greenberg}, {Sasaki}, {Bolmont}, {Grasset}, {Lewis}, {Barnes}, \& {Zuluaga}}]{heller14}
{Heller}, R., {Williams}, D., {Kipping}, D., {et~al.} 2014, \bibinfo{title}{{Formation, habitability, and detection of extrasolar moons},} Astrobiology, 14, 798, \dodoi{10.1089/ast.2014.1147}

\bibitem[{D.~C. Hsu {et~al.}(2020)Hsu, Ford, \& Terrien}]{hsu20}
Hsu, D.~C., Ford, E.~B., \& Terrien, R. 2020, \bibinfo{title}{Occurrence rates of planets orbiting M Stars: applying ABC to Kepler DR25, Gaia DR2, and 2MASS data,} \mnras, 498, 2249, \dodoi{10.1093/mnras/staa2391}

\bibitem[{O. {Jagtap} {et~al.}(2021){Jagtap}, {Quarles}, \& {Cuntz}}]{jagtap21}
{Jagtap}, O., {Quarles}, B., \& {Cuntz}, M. 2021, \bibinfo{title}{{Updated studies on exomoons in the HD 23079 system},} \pasa, 38, e059, \dodoi{10.1017/pasa.2021.52}

\bibitem[{L. {Kaltenegger}(2017){Kaltenegger}}]{kalt17}
{Kaltenegger}, L. 2017, \bibinfo{title}{{How to characterize habitable worlds and signs of life},} \araa, 55, 433, \dodoi{10.1146/annurev-astro-082214-122238}

\bibitem[{S.~R. {Kane}(2017){Kane}}]{kane17}
{Kane}, S.~R. 2017, \bibinfo{title}{{Worlds without moons: exomoon constraints for compact planetary systems},} \apjl, 839, L19, \dodoi{10.3847/2041-8213/aa6bf2}

\bibitem[{J.~F. {Kasting} {et~al.}(2014){Kasting}, {Kopparapu}, {Ramirez}, \& {Harman}}]{kasting14}
{Kasting}, J.~F., {Kopparapu}, R., {Ramirez}, R.~M., \& {Harman}, C.~E. 2014, \bibinfo{title}{{Remote life-detection criteria, habitable zone boundaries, and the frequency of Earth-like planets around M and late K stars},} Proceedings of the National Academy of Science, 111, 12641, \dodoi{10.1073/pnas.1309107110}

\bibitem[{J.~F. {Kasting} {et~al.}(1993){Kasting}, {Whitmire}, \& {Reynolds}}]{kasting93}
{Kasting}, J.~F., {Whitmire}, D.~P., \& {Reynolds}, R.~T. 1993, \bibinfo{title}{{Habitable zones around main sequence stars},} \icarus, 101, 108, \dodoi{10.1006/icar.1993.1010}

\bibitem[{D. {Kipping}(2021){Kipping}}]{kipping21}
{Kipping}, D. 2021, \bibinfo{title}{{A stationary Drake Equation distribution as a balance of birth-death processes},} Research Notes of the American Astronomical Society, 5, 44, \dodoi{10.3847/2515-5172/abeb7b}

\bibitem[{D. {Kipping} \& G. {Lewis}(2025){Kipping} \& {Lewis}}]{kipping25b}
{Kipping}, D., \& {Lewis}, G. 2025, \bibinfo{title}{{Do SETI optimists have a fine-tuning problem?},} International Journal of Astrobiology, 24, e5, \dodoi{10.1017/S1473550424000235}

\bibitem[{D. {Kipping} {et~al.}(2022){Kipping}, {Bryson}, {Burke}, {Christiansen}, {Hardegree-Ullman}, {Quarles}, {Hansen}, {Szul{\'a}gyi}, \& {Teachey}}]{kipping22}
{Kipping}, D., {Bryson}, S., {Burke}, C., {et~al.} 2022, \bibinfo{title}{{An exomoon survey of 70 cool giant exoplanets and the new candidate Kepler-1708 b-i},} Nature Astronomy, 6, 367, \dodoi{10.1038/s41550-021-01539-1}

\bibitem[{D. {Kipping} {et~al.}(2025){Kipping}, {Teachey}, {Yahalomi}, {Cassese}, {Quarles}, {Bryson}, {Hansen}, {Szul{\'a}gyi}, {Burke}, \& {Hardegree-Ullman}}]{kipping25}
{Kipping}, D., {Teachey}, A., {Yahalomi}, D.~A., {et~al.} 2025, \bibinfo{title}{{Concerning the possible exomoons around Kepler-1625 b and Kepler-1708 b},} Nature Astronomy, 9, 795, \dodoi{10.1038/s41550-025-02547-1}

\bibitem[{D.~M. {Kipping}(2009){Kipping}}]{Kipping2009}
{Kipping}, D.~M. 2009, \bibinfo{title}{{Transit timing effects due to an exomoon},} \mnras, 392, 181, \dodoi{10.1111/j.1365-2966.2008.13999.x}

\bibitem[{D.~M. {Kipping} {et~al.}(2013){Kipping}, {Forgan}, {Hartman}, {Nesvorn{\'y}}, {Bakos}, {Schmitt}, \& {Buchhave}}]{Kipping2013}
{Kipping}, D.~M., {Forgan}, D., {Hartman}, J., {et~al.} 2013, \bibinfo{title}{{The Hunt for Exomoons with Kepler (HEK). III. The First Search for an Exomoon around a Habitable-zone Planet},} \apj, 777, 134, \dodoi{10.1088/0004-637X/777/2/134}

\bibitem[{D.~M. {Kipping} {et~al.}(2014){Kipping}, {Nesvorn{\'y}}, {Buchhave}, {Hartman}, {Bakos}, \& {Schmitt}}]{hek4}
{Kipping}, D.~M., {Nesvorn{\'y}}, D., {Buchhave}, L.~A., {et~al.} 2014, \bibinfo{title}{{The hunt for exomoons with Kepler (HEK). IV. A search for moons around eight M dwarfs},} \apj, 784, 28, \dodoi{10.1088/0004-637X/784/1/28}

\bibitem[{E. {Kokubo} \& H. {Genda}(2010){Kokubo} \& {Genda}}]{Kokubo2010}
{Kokubo}, E., \& {Genda}, H. 2010, \bibinfo{title}{{Formation of terrestrial planets from protoplanets under a realistic accretion condition},} \apjl, 714, L21, \dodoi{10.1088/2041-8205/714/1/L21}

\bibitem[{R.~K. Kopparapu {et~al.}(2014)Kopparapu, Ramirez, SchottelKotte, Kasting, Domagal-Goldman, \& Eymet}]{Kopparapu_2014}
Kopparapu, R.~K., Ramirez, R.~M., SchottelKotte, J., {et~al.} 2014, \bibinfo{title}{Habitable zones around main-sequence stars: Dependence on planetary mass,} \apjl, 787, L29, \dodoi{10.1088/2041-8205/787/2/L29}

\bibitem[{R.~K. {Kopparapu} {et~al.}(2013){Kopparapu}, {Ramirez}, {Kasting}, {Eymet}, {Robinson}, {Mahadevan}, {Terrien}, {Domagal-Goldman}, {Meadows}, \& {Deshpande}}]{kopp13}
{Kopparapu}, R.~K., {Ramirez}, R., {Kasting}, J.~F., {et~al.} 2013, \bibinfo{title}{{Habitable zones around main-sequence stars: New estimates},} \apj, 765, 131, \dodoi{10.1088/0004-637X/765/2/131}

\bibitem[{P. {Kroupa}(2001){Kroupa}}]{kroupa01}
{Kroupa}, P. 2001, \bibinfo{title}{{On the variation of the initial mass function},} \mnras, 322, 231, \dodoi{10.1046/j.1365-8711.2001.04022.x}

\bibitem[{P. {Kroupa}(2002){Kroupa}}]{kroupa02}
{Kroupa}, P. 2002, \bibinfo{title}{{The initial mass function of stars: Evidence for uniformity in variable systems},} Science, 295, 82, \dodoi{10.1126/science.1067524}

\bibitem[{J. {Leconte} {et~al.}(2010){Leconte}, {Chabrier}, {Baraffe}, \& {Levrard}}]{Leconte2010}
{Leconte}, J., {Chabrier}, G., {Baraffe}, I., \& {Levrard}, B. 2010, \bibinfo{title}{{Is tidal heating sufficient to explain bloated exoplanets? Consistent calculations accounting for finite initial eccentricity},} \aap, 516, A64, \dodoi{10.1051/0004-6361/201014337}

\bibitem[{M.~A. {Limbach} {et~al.}(2024){Limbach}, {Lustig-Yaeger}, {Vanderburg}, {Vos}, {Heller}, \& {Robinson}}]{limbach24}
{Limbach}, M.~A., {Lustig-Yaeger}, J., {Vanderburg}, A., {et~al.} 2024, \bibinfo{title}{{Exomoons and exorings with the Habitable Worlds Observatory. I. On the detection of Earth{\textendash}Moon analog shadows and eclipses},} \aj, 168, 57, \dodoi{10.3847/1538-3881/ad4a75}

\bibitem[{T. {Lu} {et~al.}(2024){Lu}, {Hernandez}, \& {Rein}}]{Lu2024}
{Lu}, T., {Hernandez}, D.~M., \& {Rein}, H. 2024, \bibinfo{title}{{TRACE: a code for time-reversible astrophysical close encounters},} \mnras, 533, 3708, \dodoi{10.1093/mnras/stae1982}

\bibitem[{T. {Lu} {et~al.}(2023){Lu}, {Rein}, {Tamayo}, {Hadden}, {Mardling}, {Millholland}, \& {Laughlin}}]{tides_spin}
{Lu}, T., {Rein}, H., {Tamayo}, D., {et~al.} 2023, \bibinfo{title}{{Self-consistent spin, tidal, and dynamical equations of motion in the REBOUNDx framework},} \apj, 948, 41, \dodoi{10.3847/1538-4357/acc06d}

\bibitem[{R. {Luger} \& R. {Barnes}(2015){Luger} \& {Barnes}}]{luger15}
{Luger}, R., \& {Barnes}, R. 2015, \bibinfo{title}{{Extreme water loss and abiotic O2 buildup on planets throughout the habitable zones of M dwarfs},} Astrobiology, 15, 119, \dodoi{10.1089/ast.2014.1231}

\bibitem[{T. {LUVOIR Team}(2019){LUVOIR Team}}]{luvo19}
{LUVOIR Team}, T. 2019, The LUVOIR Mission concept study final report, \doarXiv{1912.06219}

\bibitem[{H. Martínez-Rodríguez {et~al.}(2019)Martínez-Rodríguez, Caballero, Cifuentes, Piro, \& Barnes}]{marrod19}
Martínez-Rodríguez, H., Caballero, J.~A., Cifuentes, C., Piro, A.~L., \& Barnes, R. 2019, \bibinfo{title}{Exomoons in the habitable zones of M dwarfs,} \apj, 887, 261, \dodoi{10.3847/1538-4357/ab5640}

\bibitem[{S.~C. Millholland \& J.~N. Winn(2025)Millholland \& Winn}]{Millholland2025}
Millholland, S.~C., \& Winn, J.~N. 2025, \bibinfo{title}{Exploring exoplanet dynamics with JWST: Tides, rotation, rings, and moons,} Proceedings of the National Academy of Sciences, 122, e2416189122, \dodoi{10.1073/pnas.2416189122}

\bibitem[{G.~D. {Mulders} {et~al.}(2015){Mulders}, {Pascucci}, \& {Apai}}]{mulders15}
{Mulders}, G.~D., {Pascucci}, I., \& {Apai}, D. 2015, \bibinfo{title}{{An increase in the mass of planetary systems around lower-mass stars},} \apj, 814, 130, \dodoi{10.1088/0004-637X/814/2/130}

\bibitem[{E. {National Academies of Sciences}(2021){National Academies of Sciences}}]{HWO}
{National Academies of Sciences}, E. 2021, {Pathways to discovery in astronomy and astrophysics for the 2020s}, \dodoi{10.17226/26141}

\bibitem[{O. {Neron de Surgy} \& J. {Laskar}(1997){Neron de Surgy} \& {Laskar}}]{neron97}
{Neron de Surgy}, O., \& {Laskar}, J. 1997, \bibinfo{title}{{On the long term evolution of the spin of the Earth.},} \aap, 318, 975

\bibitem[{J.~P. {Noyola} {et~al.}(2014){Noyola}, {Satyal}, \& {Musielak}}]{noyola14}
{Noyola}, J.~P., {Satyal}, S., \& {Musielak}, Z.~E. 2014, \bibinfo{title}{{Detection of exomoons through observation of radio emissions},} \apj, 791, 25, \dodoi{10.1088/0004-637X/791/1/25}

\bibitem[{J.~P. {Noyola} {et~al.}(2016){Noyola}, {Satyal}, \& {Musielak}}]{noyola16}
{Noyola}, J.~P., {Satyal}, S., \& {Musielak}, Z.~E. 2016, \bibinfo{title}{{On the radio detection of multiple-exomoon systems due to plasma torus sharing},} \apj, 821, 97, \dodoi{10.3847/0004-637X/821/2/97}

\bibitem[{E. {Pass} {et~al.}(2024){Pass}, {Bean}, {Charbonneau}, {Cherubim}, \& {Garcia-Mejia}}]{jwst24}
{Pass}, E., {Bean}, J.~L., {Charbonneau}, D., {Cherubim}, C., \& {Garcia-Mejia}, J. 2024, {A search for exoplanet satellites that are the same size as the Earth's moon},, JWST Proposal. Cycle 3, ID. \#6193

\bibitem[{S.~D. {Patel} {et~al.}(2025){Patel}, {Quarles}, \& {Cuntz}}]{patel25}
{Patel}, S.~D., {Quarles}, B., \& {Cuntz}, M. 2025, \bibinfo{title}{{Orbital stability of hierarchical three- and four-body systems with inclination: results for Kepler-1625, 1708, and HD 23079},} \mnras, 537, 2291, \dodoi{10.1093/mnras/staf131}

\bibitem[{S.~D. Patel {et~al.}(2025)Patel, Quarles, Cuntz, \& Weinberg}]{patel25b}
Patel, S.~D., Quarles, B., Cuntz, M., \& Weinberg, N.~N. 2025, \bibinfo{title}{Can moons exist around the habitable-zone planet K2-18b?} Monthly Notices of the Royal Astronomical Society: Letters, 542, L144, \dodoi{10.1093/mnrasl/slaf077}

\bibitem[{M.~J. Pecaut \& E.~E. Mamajek(2013)Pecaut \& Mamajek}]{Pecaut_2013}
Pecaut, M.~J., \& Mamajek, E.~E. 2013, \bibinfo{title}{Intrinsic colors, temperatures, and bolometric corrections of pre-main-sequence stars,} \apjs, 208, 9, \dodoi{10.1088/0067-0049/208/1/9}

\bibitem[{A.~L. {Piro}(2018){Piro}}]{piro18}
{Piro}, A.~L. 2018, \bibinfo{title}{{Exoplanets torqued by the combined tides of a moon and parent star},} \aj, 156, 54, \dodoi{10.3847/1538-3881/aaca38}

\bibitem[{R.~M. {Ramirez}(2018){Ramirez}}]{ram18}
{Ramirez}, R.~M. 2018, \bibinfo{title}{{A more comprehensive habitable zone for finding life on other planets},} Geosciences, 8, 280, \dodoi{10.3390/geosciences8080280}

\bibitem[{R.~M. {Ramirez} \& L. {Kaltenegger}(2014){Ramirez} \& {Kaltenegger}}]{ram14}
{Ramirez}, R.~M., \& {Kaltenegger}, L. 2014, \bibinfo{title}{{The habitable zones of pre-main-sequence stars},} \apjl, 797, L25, \dodoi{10.1088/2041-8205/797/2/L25}

\bibitem[{H. {Rein} \& S.~F. {Liu}(2012){Rein} \& {Liu}}]{rebound}
{Rein}, H., \& {Liu}, S.~F. 2012, \bibinfo{title}{{REBOUND: an open-source multi-purpose N-body code for collisional dynamics},} \aap, 537, A128, \dodoi{10.1051/0004-6361/201118085}

\bibitem[{H. {Rein} \& D.~S. {Spiegel}(2015){Rein} \& {Spiegel}}]{Rein2015}
{Rein}, H., \& {Spiegel}, D.~S. 2015, \bibinfo{title}{{IAS15: a fast, adaptive, high-order integrator for gravitational dynamics, accurate to machine precision over a billion orbits},} \mnras, 446, 1424, \dodoi{10.1093/mnras/stu2164}

\bibitem[{M. {Rosario-Franco} {et~al.}(2020){Rosario-Franco}, {Quarles}, {Musielak}, \& {Cuntz}}]{Rosario-Franco2020}
{Rosario-Franco}, M., {Quarles}, B., {Musielak}, Z.~E., \& {Cuntz}, M. 2020, \bibinfo{title}{{Orbital Stability of exomoons and submoons with applications to Kepler 1625b-I},} \aj, 159, 260, \dodoi{10.3847/1538-3881/ab89a7}

\bibitem[{T. {Sasaki} {et~al.}(2012){Sasaki}, {Barnes}, \& {O'Brien}}]{sasaki12}
{Sasaki}, T., {Barnes}, J.~W., \& {O'Brien}, D.~P. 2012, \bibinfo{title}{{Outcomes and duration of tidal evolution in a star-planet-moon system},} \apj, 754, 51, \dodoi{10.1088/0004-637X/754/1/51}

\bibitem[{Y. {Su} \& M. {Saillenfest}(2025){Su} \& {Saillenfest}}]{Su2025}
{Su}, Y., \& {Saillenfest}, M. 2025, \bibinfo{title}{{The Coupled Tidal Evolution of the Moons and Spins of Warm Exoplanets},} arXiv e-prints, arXiv:2511.00161.
\newblock \doarXiv{2511.00161}

\bibitem[{W.~T. {Sullivan} \& J. {Baross}(2001){Sullivan} \& {Baross}}]{sullivan01}
{Sullivan}, III, W.~T., \& {Baross}, J. 2001, {Planets and Life} (Cambridge University Press)

\bibitem[{G.~M. {Szab{\'o}} {et~al.}(2024){Szab{\'o}}, {Schneider}, {Dencs}, \& {K{\'a}lm{\'a}n}}]{szabo24}
{Szab{\'o}}, G.~M., {Schneider}, J., {Dencs}, Z., \& {K{\'a}lm{\'a}n}, S. 2024, \bibinfo{title}{{The 'Drake Equation' of exomoons{\textemdash}A cascade of formation, stability and detection},} Universe, 10, 110, \dodoi{10.3390/universe10030110}

\bibitem[{K. {Takaoka} {et~al.}(2023){Takaoka}, {Kuwahara}, {Ida}, \& {Kurokawa}}]{Takaoka2023}
{Takaoka}, K., {Kuwahara}, A., {Ida}, S., \& {Kurokawa}, H. 2023, \bibinfo{title}{{Spin of protoplanets generated by pebble accretion: Influences of protoplanet-induced gas flow},} \aap, 674, A193, \dodoi{10.1051/0004-6361/202345915}

\bibitem[{D. {Tamayo} {et~al.}(2020){Tamayo}, {Rein}, {Shi}, \& {Hernandez}}]{reboundx}
{Tamayo}, D., {Rein}, H., {Shi}, P., \& {Hernandez}, D.~M. 2020, \bibinfo{title}{{REBOUNDx: a library for adding conservative and dissipative forces to otherwise symplectic N-body integrations},} \mnras, 491, 2885, \dodoi{10.1093/mnras/stz2870}

\bibitem[{A. {Teachey}(2024){Teachey}}]{teachey2024}
{Teachey}, A. 2024, \bibinfo{title}{{Detecting and characterizing exomoons and exorings (Handbook of Exoplanets, 2nd Edition)},} arXiv e-prints, arXiv:2401.13293, \dodoi{10.48550/arXiv.2401.13293}

\bibitem[{A. {Teachey} \& D.~M. {Kipping}(2018){Teachey} \& {Kipping}}]{teachey18}
{Teachey}, A., \& {Kipping}, D.~M. 2018, \bibinfo{title}{{Evidence for a large exomoon orbiting Kepler-1625b},} Science Advances, 4, eaav1784, \dodoi{10.1126/sciadv.aav1784}

\bibitem[{C.~G. {Tinney} {et~al.}(2002){Tinney}, {Butler}, {Marcy}, {Jones}, {Penny}, {McCarthy}, \& {Carter}}]{tinney02}
{Tinney}, C.~G., {Butler}, R.~P., {Marcy}, G.~W., {et~al.} 2002, \bibinfo{title}{{Two extrasolar planets from the Anglo-Australian Planet Search},} \apj, 571, 528, \dodoi{10.1086/339916}

\bibitem[{T. {Trifonov} {et~al.}(2020){Trifonov}, {Lee}, {K{\"u}rster}, {Henning}, {Grishin}, {Stock}, {Tjoa}, {Caballero}, {Wong}, {Bauer}, {Quirrenbach}, {Zechmeister}, {Ribas}, {Reffert}, {Reiners}, {Amado}, {Kossakowski}, {Azzaro}, {B{\'e}jar}, {Cort{\'e}s-Contreras}, {Dreizler}, {Hatzes}, {Jeffers}, {Kaminski}, {Lafarga}, {Montes}, {Morales}, {Pavlov}, {Rodr{\'\i}guez-L{\'o}pez}, {Schmitt}, {Solano}, \& {Barnes}}]{trif20}
{Trifonov}, T., {Lee}, M.~H., {K{\"u}rster}, M., {et~al.} 2020, \bibinfo{title}{{The CARMENES search for exoplanets around M dwarfs. Dynamical characterization of the multiple planet system GJ 1148 and prospects of habitable exomoons around GJ 1148 b},} \aap, 638, A16, \dodoi{10.1051/0004-6361/201936987}

\bibitem[{A. {Vanderburg} {et~al.}(2018){Vanderburg}, {Rappaport}, \& {Mayo}}]{vanderburg18}
{Vanderburg}, A., {Rappaport}, S.~A., \& {Mayo}, A.~W. 2018, \bibinfo{title}{{Detecting exomoons via Doppler monitoring of directly imaged exoplanets},} \aj, 156, 184, \dodoi{10.3847/1538-3881/aae0fc}

\bibitem[{Y.-T. {Wang} {et~al.}(2025){Wang}, {Liu}, \& {Li}}]{wang25}
{Wang}, Y.-T., {Liu}, C., \& {Li}, J. 2025, \bibinfo{title}{{Stellar initial mass function in the 100-pc solar neighbourhood},} arXiv e-prints, arXiv:2506.12987, \dodoi{10.48550/arXiv.2506.12987}

\bibitem[{D.~M. {Williams} {et~al.}(1997){Williams}, {Kasting}, \& {Wade}}]{williams97}
{Williams}, D.~M., {Kasting}, J.~F., \& {Wade}, R.~A. 1997, \bibinfo{title}{{Habitable moons around extrasolar giant planets},} \nat, 385, 234, \dodoi{10.1038/385234a0}

\bibitem[{R.~R. {Zollinger} {et~al.}(2017){Zollinger}, {Armstrong}, \& {Heller}}]{zoll17}
{Zollinger}, R.~R., {Armstrong}, J.~C., \& {Heller}, R. 2017, \bibinfo{title}{{Exomoon habitability and tidal evolution in low-mass star systems},} \mnras, 472, 8, \dodoi{10.1093/mnras/stx1861}

\end{thebibliography}
\bibliographystyle{aasjournalv7}

\end{document}